\documentclass[journal]{IEEEtran}
\usepackage{xcolor}
\usepackage{lineno,hyperref}
\usepackage{amsmath}
\usepackage{mathtools}
\usepackage{float}
\usepackage[caption=false]{subfig}
\usepackage{csquotes}
\usepackage{color,soul}
\usepackage[shortlabels]{enumitem}
\usepackage{multirow}
\usepackage{amsfonts}
\usepackage{amssymb}
\usepackage{cite}
\usepackage{tabularx}
\usepackage{adjustbox}
\usepackage{bm}
\usepackage{blindtext, rotating}

\begin{document}
\title{CG-Net: Conditional GIS-aware Network for Individual Building Segmentation in VHR SAR Images}
\author{Yao~Sun,
        Yuansheng~Hua, 
        Lichao~Mou, 
        and~Xiao~Xiang~Zhu,~\IEEEmembership{Senior~Member,~IEEE}
\thanks{This work is jointly supported by the European Research Council (ERC) under the European Union's Horizon 2020 research and innovation programme (grant agreement No. [ERC-2016-StG-714087], Acronym: \textit{So2Sat}), by the Helmholtz Association
through the Framework of Helmholtz Artificial Intelligence Cooperation Unit (HAICU) - Local Unit ``Munich Unit @Aeronautics, Space and Transport (MASTr)'' and Helmholtz Excellent Professorship ``Data Science in Earth Observation - Big Data Fusion for Urban Research'', and by the German Federal Ministry of Education and Research (BMBF) in the framework of the international future AI lab ``AI4EO -- Artificial Intelligence for Earth Observation: Reasoning, Uncertainties, Ethics and Beyond''.}
\thanks{Y. Sun, Y. Hua, L. Mou, and X. X. Zhu are with the Remote
Sensing Technology Institute, German Aerospace Center, 82234 Wessling, Germany, and also with the Signal Processing in Earth Observation, Technical
University of Munich, 80333 Munich, Germany. (e-mails: yao.sun@dlr.de; yuansheng.hua@dlr.de; lichao.mou@dlr.de; xiaoxiang.zhu@dlr.de)}
}

\markboth{Accepted article: IEEE Transactions on Geoscience and Remote Sensing, November~2020}%
{Shell \MakeLowercase{\textit{et al.}}: Bare Demo of IEEEtran.cls for Journals}

\maketitle

\begin{abstract}
Object retrieval and reconstruction from very high resolution (VHR) synthetic aperture radar (SAR) images are of great importance for urban SAR applications, yet highly challenging owing to the complexity of SAR data. 
This paper addresses the issue of individual building segmentation from a single VHR SAR image in large-scale urban areas. 
To achieve this, we introduce building footprints from GIS data as complementary information and propose a novel conditional GIS-aware network (CG-Net). The proposed model learns multi-level visual features and employs building footprints to normalize the features for predicting building masks in the SAR image. 
We validate our method using a high resolution spotlight TerraSAR-X image collected over Berlin. Experimental results show that the proposed CG-Net effectively brings improvements with variant backbones. 
We further compare two representations of building footprints, namely complete building footprints and sensor-visible footprint segments, for our task, and conclude that the use of the former leads to better segmentation results. 
Moreover, we investigate the impact of inaccurate GIS data on our CG-Net, and this study shows that CG-Net is robust against  positioning errors in GIS data. 
In addition, we propose an approach of ground truth generation of buildings from an accurate digital elevation model (DEM), which can be used to generate large-scale SAR image datasets. 
The segmentation results can be applied to reconstruct 3D building models at level-of-detail (LoD) 1, which is demonstrated in our experiments.
\end{abstract}


\IEEEpeerreviewmaketitle

    
\section{Introduction}\label{sec:intro}

\IEEEPARstart{V}{ery} high resolution (VHR) synthetic aperture radar (SAR) imagery has attracted many researchers in modeling and characterization of objects of interest in urban environments \cite{franceschetti2002Canonical,tupin2003Detection,guida2010Height,brunner2010Building,sportouche2011Extraction,wenliu2013Building,zhu2014Facade}, as it is able to provide data being independent of sun illumination and insensitive to weather conditions. 
Such data source is particularly of interest to studies concerning areas frequently covered by clouds \cite{huang2015Cloud} and to applications of emergency response \cite{brunner2010Earthquake,wang2012Postearthquake}. 
However, because of side-looking imaging geometry and complex backscattering mechanism, SAR image interpretation is challenging, especially in urban areas where severe geometric distortions such as layover and shadowing further complicate SAR image understanding. 

     \begin{figure}[!]
        \centering
        \subfloat{\includegraphics[width= 0.331\columnwidth]{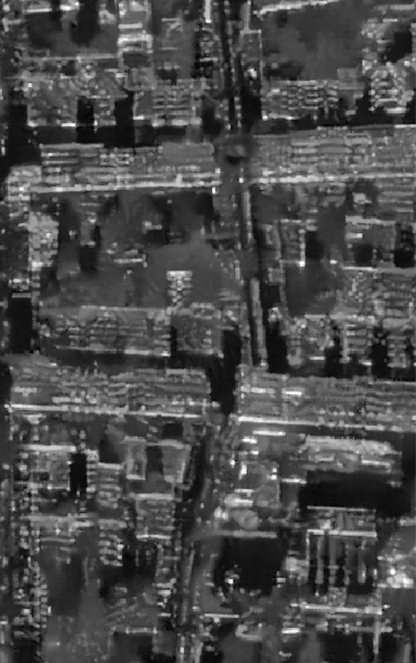}}\hfill
        \subfloat{\includegraphics[width= 0.331\columnwidth]{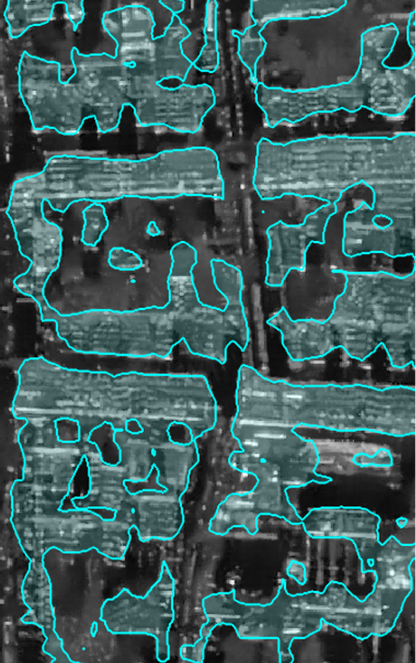}}\hfill
        \subfloat{\includegraphics[width= 0.331\columnwidth]{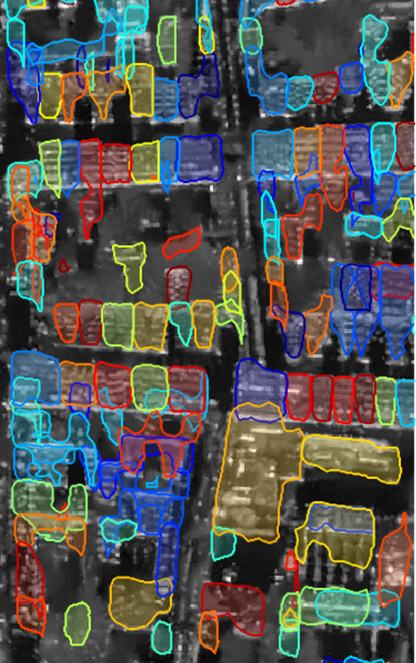}}
        \caption{Illustration of the difference between building semantic segmentation and individual building segmentation. From left to right: a SAR image, the result of building semantic segmentation \cite{sun2019Largescale}, and the result of individual building segmentation (ours). In the middle image, all buildings are assigned the same label, while in the right image, each individual building is identified as one class.
        }
        \label{fig:demo}
    \end{figure}
    
Buildings are the dominant structures in urban regions. 
The literature on retrieving information (e.g., footprint and height) from individual buildings on a large-scale VHR SAR image is still in its infancy. 
In \cite{shahzad2019Buildings, sun2019Largescale}, buildings are segmented from large-scale SAR images using deep networks. 
However, individual buildings cannot be recognized, due to 
serious layover effects on high-rise buildings in urban areas. 
Fig. \ref{fig:demo} shows the difference between building semantic segmentation results (middle) and our individual building segmentation results (right) in a SAR image (left). As can be seen, the latter is capable of not only providing pixel-wise segmentation masks but also separating building instances. 
On the other hand, 
several works \cite{guida2010Height, brunner2010Building, sportouche2011Extraction} develop tailored algorithms to perform accurate analyses for buildings in complex urban environments, but these methods are limited to be applied for large-scale areas. 
In this work, we are interested in individual building segmentation from SAR images in a large scale. 
In what follows, we briefly explain challenges of this task and review related work.

\subsection{Challenges}
    
    Interpreting individual buildings in SAR images is highly challenging, mainly for two reasons. 
    First, intensity values in SAR images are closely related to material types and structural shapes of objects. 
    Therefore, consecutive buildings in the physical world are difficult to be separated from each other in a SAR image, unless in the presence of obvious material or structure changes at building boundaries. 
    Second, even if buildings in the real world are not neighboring, they probably overlap with each other in the SAR image, which significantly increases the difficulty of image interpretation. 
    Fig. \ref{fig:samples} shows two typical urban areas in an optical image (the first column) and a VHR SAR image (the second column). Footprints and regions of buildings present in the SAR image are marked with different colors as shown in the following two columns.
    It can be seen that some buildings severely overlap in the SAR image even if their corresponding footprints are not next to each other.

\subsection{Related Work}
 
Generally, building extraction approaches from SAR data can be grouped into the following two categories: data-driven methods and model-driven methods.  
The former extracts building features and then deduces building parameters. Two solutions based on this methodology have been developed. 
The first one makes an attempt at detecting line- or point-like features first and extracting building regions based on these features. 
For example,  
in \cite{tupin2003Detection}, feature lines are identified using a line detector, and layover areas are derived by extracting parallel edges;  
in \cite{xu2007Automatic}, the authors exploit a constant false alarm rate (CFAR) edge detector for line feature detection and apply a Hough transform for parallelogram-like wall area extraction; 
in \cite{michaelsen2006Perceptual,soergel2009Stereo}, bright line segments and regular spaced point-like features are detected and subsequently grouped to building footprints; 
and in \cite{ferro2013Automatic}, the authors extract and combine a set of low-level features to create structured primitives.  
The second solution directly extracts building regions using segmentation techniques, such as active contour \cite{hill2008Estimating}, rotating mask \cite{bolterShapefromShadow}, mean-shift \cite{cellier2005introduction}, and marker-controlled watershed transform \cite{zhao2013Building}.
In model-driven methods, a SAR image or InSAR phase is iteratively simulated using geometric and radiometric hypothesis \cite{sportouche2009Building, brunner2010Building,thiele2012gis,zhang2015Stochastic,  quartulli2004Stochastic, guida2008ModelBased, guida2010Height}. The desired building parameters are progressively achieved by minimizing the difference between simulated and real data. 

\newcolumntype{Y}{>{\centering\arraybackslash}X}

    \begin{figure}[!]
        \centering
        \subfloat{\includegraphics[width=.247\columnwidth]{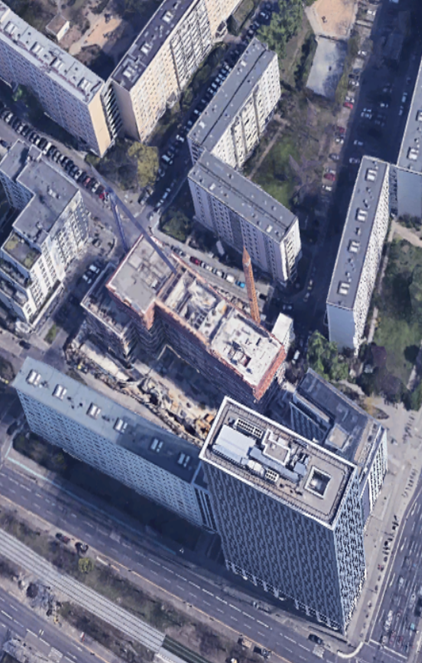}}\hfill
        \subfloat{\includegraphics[width=.247\columnwidth]{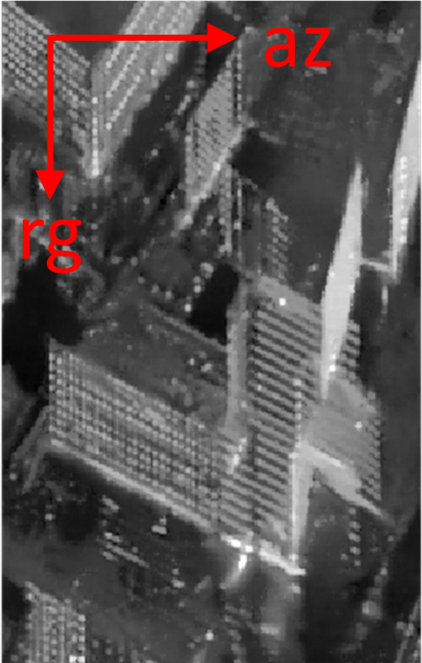}}\hfill
        \subfloat{\includegraphics[width=.247\columnwidth]{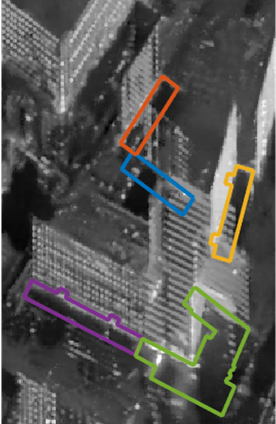}}\hfill
        \subfloat{\includegraphics[width=.247\columnwidth]{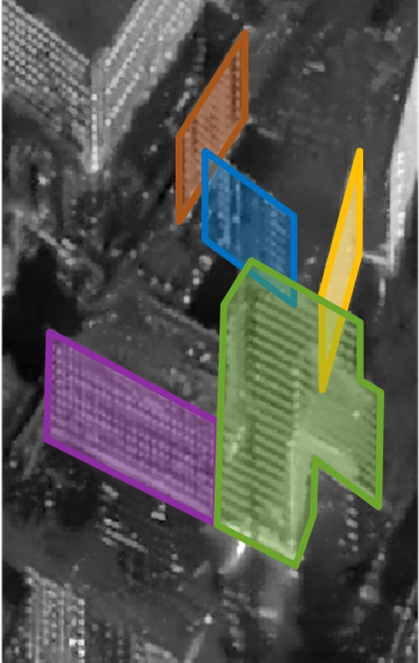}}\\
        \vspace{-0.3cm}
        \subfloat{\includegraphics[width=.247\columnwidth]{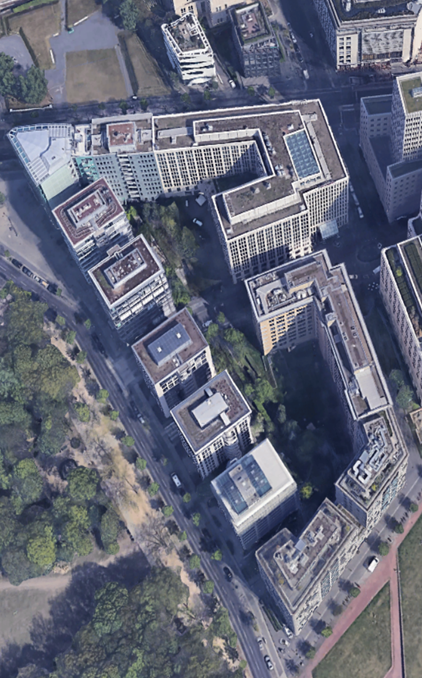}}\hfill
        \subfloat{\includegraphics[width=.247\columnwidth]{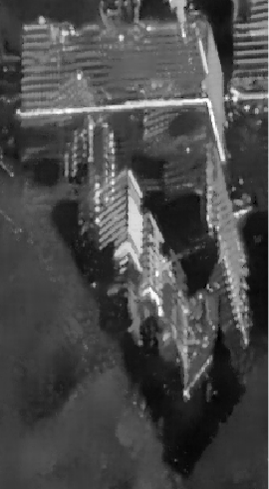}}\hfill
        \subfloat{\includegraphics[width=.247\columnwidth]{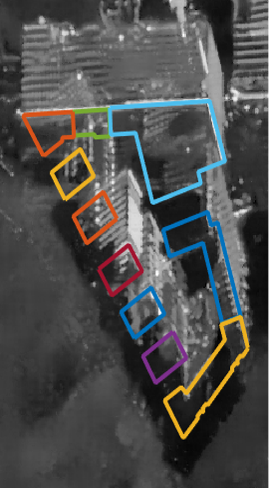}}\hfill
        \subfloat{\includegraphics[width=.247\columnwidth]{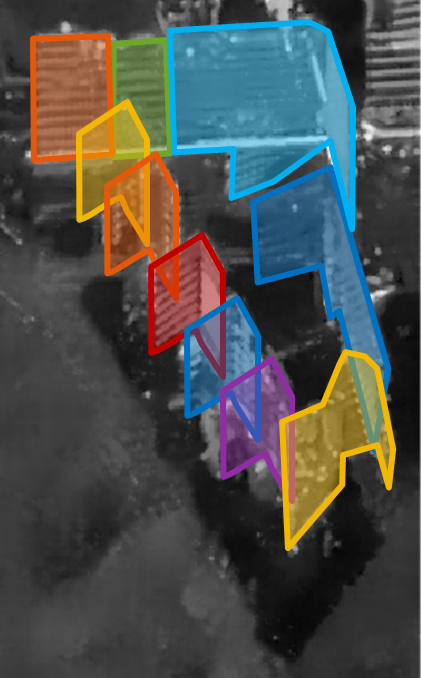}}\\
        \begin{tabularx}{\columnwidth}{@{}lYYYY@{}}
        {\hskip6pt}\textbf{\footnotesize{Optical image}} & {\hskip3pt}\textbf{\footnotesize SAR image}  & \textbf{\footnotesize Footprints} &  \textbf{\footnotesize Buildings}
        \end{tabularx}
        \vspace{-0.3cm}
        \caption{Two typical urban areas shown in an optical image and a SAR image. 
        In column 3 and 4, footprints and the corresponding building regions in the SAR image are marked in different colors for reference.  
        rg and az denote the range direction and azimuth direction, respectively.}
        \label{fig:samples}
    \end{figure}    

The majority of related studies are carried out on buildings with specific geometric shapes, e.g., rectangular- \cite{simonetto2005Rectangular, wang2008Building,liu2017BottomUp} or L-shaped footprints \cite{zhang2011Building, zhao2013Building}, flat \cite{wegner2014Combining} or gable roofs \cite{thiele2010Analysis, chen2017Automatic}, and different heights \cite{chen2017Automatic,guo2014HighRise,liu2015Height,tang2016Highrise}.
Only a few studies address the problem of complex-shaped buildings \cite{michaelsen2006Perceptual,soergel2009Stereo}.  
Furthermore, most studies investigate simple scenarios where a minimal distance between buildings is required to ensure scattering effects of different buildings do not interfere with each other \cite{guida2010Height, brunner2010Building, sportouche2011Extraction, shanshanchen2015Automatic}. 
In complex scenarios, possible overlapping areas between two buildings are usually assigned to one building \cite{wenliu2013Building,lu2015New}, which may cause incorrect estimations. 
By using a SAR tomography (TomoSAR) point cloud, Shahzad \textit{et al.} \cite{shahzad2014Reconstructing}  extract buildings without imposing constraints on building shapes and study scenarios. However, the TomoSAR technique \cite{zhu2010Very} requires multiple SAR acquisitions that are generally unavailable for most areas and for applications with a stringent time limit, such as emergency response. 

In addition to SAR data, some auxiliary data are introduced, e.g., building outlines extracted from optical images \cite{wegner2009Building,sportouche2011Extraction} and footprint polygons obtained from GIS data \cite{thiele2010Combining,wenliu2013Building,yaosun2017Building}, for providing exact locations and geometric shapes of buildings in the real world. 
As illustrated in Fig. \ref{fig:samples}, in complex urban regions, the use of footprints is beneficial for tasks concerning individual buildings in SAR images. 
In exploiting the shape information, sensor-visible footprint segments, i.e., near-range segments in footprint polygons that correspond to sensor-visible walls, are desirable for extracting layover areas \cite{wenliu2013Building, yaosun2017Building}; contrarily, complete building footprints may provide additional information especially for extracting roof areas of low-rise buildings \cite{sportouche2011Extraction}. Therefore, it leaves a question on how footprints can be effectively used. We demonstrate this issue in this work by comparing results from both the footprint utilizations. 

        \begin{figure*}[!]
            \centering
            \includegraphics[width = \linewidth]{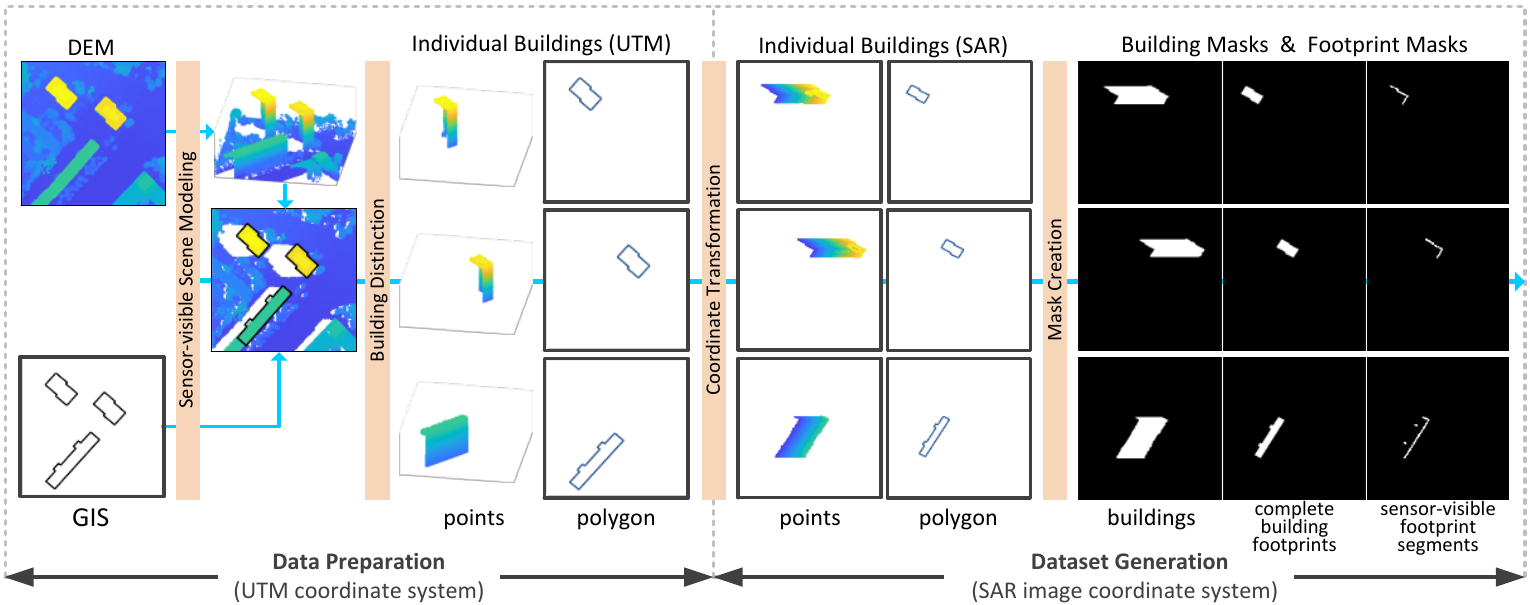}
            \caption{The workflow for dataset generation. We first collect DEM and GIS data in the UTM coordinate system and then project them to the SAR image coordinate system in order to generate building ground truth annotations and the corresponding footprints in our study area.}
            \label{fig:datawf}
        \end{figure*}

In recent years, deep neural networks have been becoming increasingly popular and shown success in remote-sensing data analysis \cite{zhang2016deep, 8113128,mou2019relation,kussul2017deep,cheng2018deep,mou2018im2height,kemker2018algorithms,audebert2018beyond,hua2020relation,mou2020relation}, including a wide range of  applications using SAR data, such as classification \cite{lv2015Urban,zhou2016Polarimetric,zhang2017ComplexValued,zhao2017discriminant,geng2017Deep}, segmentation \cite{duan_sar_2017,mohammadimanesh2019New}, target recognition \cite{chen2014SAR,ding2016Convolutional,kang2017contextual,gao2019New}, and change detection \cite{gong2016Change, gao2016Automatic, geng2019SaliencyGuided}. Instead of relying on hand-crafted features, deep networks can learn effective feature representations from raw data in an end-to-end fashion. But one problem of applying deep networks to urban SAR analysis tasks is the lack of annotation data. 
To address this issue, Wang \textit{et al.} \cite{wang2019HRSARNet} take building polygons from the OpenStreetMap (OSM) dataset and an official map as ground truth data and train a network 
to segment buildings in an urban scene. 
For building footprint extraction, Shermeyer \textit{et al.} \cite{shermeyer2020spacenet} present a multi-sensor all weather mapping (MSAW) dataset containing airborne SAR images, optical images, and building footprint annotations, along with a deep network baseline model and benchmark.  
However, in these two works, building footprints, instead of building areas, are learning targets. 
By introducing a TomoSAR point cloud, Shahzad \textit{et al.} \cite{shahzad2019Buildings} are able to acquire accurate building areas in a SAR image and take them as ground truth annotations to train a segmentation network for the purpose of building extraction. 
However, this work cannot differentiate individual buildings. 
As our survey of related work shows above, there is a paucity of literature on using deep learning for VHR SAR image interpretation in complex urban areas, particularly aiming at segmenting individual and overlapping buildings.

    \subsection{Contributions}
    
    In this work, we intend to segment individual buildings in a large urban area by exploiting SAR images and building footprints. 
    For the training of models, we generate pixel-wise ground truth annotations from an accurate DEM. 
    And building footprints are acquired from GIS data.
    Afterwards, a novel conditional GIS-aware network (CG-Net) has been proposed to first learn multi-level visual features and then employ GIS building footprint data to normalize these features for predicting final building masks.  
    In addition, we compare two representations of building footprints, namely complete building footprints and sensor-visible footprint segments, aiming to find out a more suitable representation way for this task. 
%
%
The main contributions of this paper are in four-fold:
\begin{itemize}
    \item
    We propose a workflow for the segmentation of individual buildings in VHR SAR images with GIS data. 
    To our best knowledge, this is the first time that individual buildings are studied on a large-scale SAR image, 
    and deep networks are employed in the problem of individual building segmentation of SAR images. 

    \item 
    We propose a network termed as CG-Net, which is capable of significantly improving the performance of networks for our task by imposing constraints on the learning process. 
    \item 
    We investigate the impact of inaccurate GIS data on CG-Net and find out that CG-Net is robust against positioning errors in GIS data. This study suggests that the large amount of open sourced GIS data can be exploited for individual building segmentation in SAR images.
    \item 
    We propose a ground truth generation approach to produce building masks using an accurate DEM. We believe that our method can provide large potential in analyzing complex urban regions. 
    
\end{itemize}

    The remainder of this paper is organized as follows. 
    The detailed procedure of the dataset generation is presented in Section \ref{sec:data}, and the proposed CG-Net is delineated in Section \ref{sec:CG}. 
    Section \ref{sec:test} introduces the configuration of experiments and analyzes results. Section \ref{sec:app} demonstrates an application using the produced segmentation results. In Section \ref{sec:conclude}, we conclude this paper.

\section{Dataset generation}\label{sec:data}

\subsection{Overview}

    Building annotations (as ground truth data) and building footprints (as input data) in SAR images are necessary for training our network.   
    For this reason, we propose a workflow that employs a highly accurate DEM and GIS building footprints to automatically label building masks and their corresponding footprints in SAR images. 
    Our dataset is generated in two stages. First, sensor-visible 3D building models (i.e., non-occluded roofs and facades) and building footprints are prepared in the UTM coordinate system. 
    Second, they are projected to the SAR image coordinate system in order to generate building ground truth annotations and the corresponding footprints. 
    Fig. \ref{fig:datawf} illustrates the workflow, and for more details, refer to the following sections. 
    
        \begin{figure}[!]
            \centering
            \includegraphics[width = 0.9 \columnwidth]{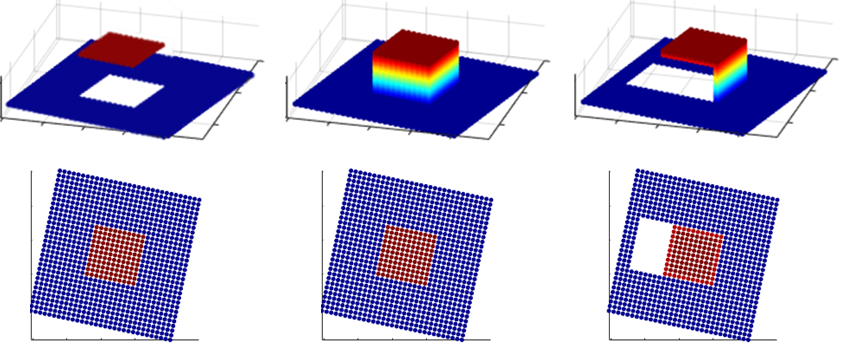}
            \begin{tabular}{ccc}
            $P_{dem}$ \hspace{1.3cm} & $P_{com} $ \hspace{1.3cm} & $P_{svs}$  \\
            \end{tabular}
            \caption{Illustration of scene modeling steps with a simulated DEM in 3D (first row) and 2D (second row): (left) the DEM point cloud $P_{dem}$; (middle) the complete point cloud $P_{com}$ after adding vertical points; (right) the sensor-visible point cloud $P_{svs}$ after hidden point removal. }
            \label{fig:hpr}
        \end{figure}

         \begin{figure}[!]
            \centering
            \includegraphics[trim=0cm 1cm 0cm 0cm, clip=true,width=.95\columnwidth]{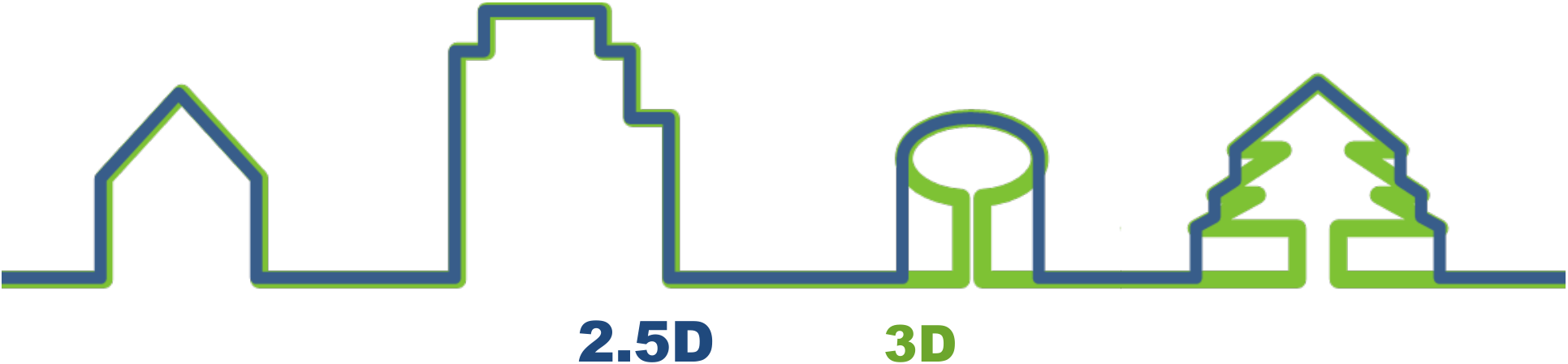}
            \\
            \hspace{-0.76cm}\footnotesize{\textbf{buildings}} \hspace{3cm}  \footnotesize{\textbf{trees}}
            \caption{Illustration of 2.5D (dark blue) and 3D (green) surface models. In 2.5D representation, each 2D point $(x, y)$ is assigned to a unique height value $z$. Therefore 2.5D DEM can represent vertical walls of buildings, but not vertical surfaces of complex objects, such as trees. }
            \label{fig:3d}
        \end{figure} 

\subsection{Data Preparation in the UTM Coordinate System}

    \subsubsection{Sensor-visible Scene Modeling}
    We first model a scene that can be viewed by a radar sensor in the UTM coordinate system. The procedure is conducted in three steps (cf. Fig. \ref{fig:hpr}):
        \begin{itemize}
        \item
        \textit{DEM is transformed to a point cloud $P_{dem}$}.
        Specifically, each pixel in the DEM with geolocation coordinates $(x,y)$ and a height value $h$ is represented as a point with coordinates $(x,y,h)$, and hence all pixels establish a nadir-looking 3D point cloud $P_{dem}$.
        \item
        \textit{A complete 3D point cloud $P_{com}$ is generated by filling vertical data gaps}. To be more specific, vertical structures such as building walls that are absent from $P_{dem}$ are added through the following steps. 
        We first detect building points which are located at height jumps. 
        Afterwards, at each detected point $g(x,y,h)$, 
        a vertical point set $G = \{{g_i(x_i,y_i,h_i)| i=1,...,m}\}$ is added, where $ x_i=x, y_i=y, h_{i} = h_0 + i \times h_{step}, h_{i}<h_e.$  
        $h_0$ and $h_e$ are the minimum and maximum heights in the neighbourhood of $g$, $h_{step}$ is a predefined height step, and the number of points $m=(h_e - h_0)/h_{step}$. 
        Eventually, a complete 3D point cloud $P_{com}$ is built by all vertical point sets and $P_{dem}$. 
        Note that the DEM is 2.5D instead of true 3D, i.e., each 2D point $(x, y)$ is assigned to a unique height value $z$ \cite{weibel1991digital}, that vertical surfaces of complex objects are not represented, such as trees (cf. Fig. \ref{fig:3d}). Therefore vertical points are only added to building areas in this step.

        \item
        \textit{A sensor-visible scene point cloud $P_{svs}$ is obtained through a visibility test on the point cloud $P_{com}$.}  Since a radar sensor only sees one side of a scene, points on the other side should be removed. 
        To this end, the hidden point removal (HPR) algorithm \cite{katz2007Direct} is applied. 
       
       In our process, the viewpoint in HPR is positioned on the line of sight of the radar sensor at a large distance away from the scene, in order to simulate an orthographic view in the azimuth of the radar sensor. 
       In this way, sight lines from the viewpoint to objects in the scene are parallel to each other and orthogonal to the azimuth, enabling HPR to remove sensor-invisible points.
       
    \end{itemize}

        \begin{figure}[!]
            \centering
            \includegraphics[width=.95\columnwidth]{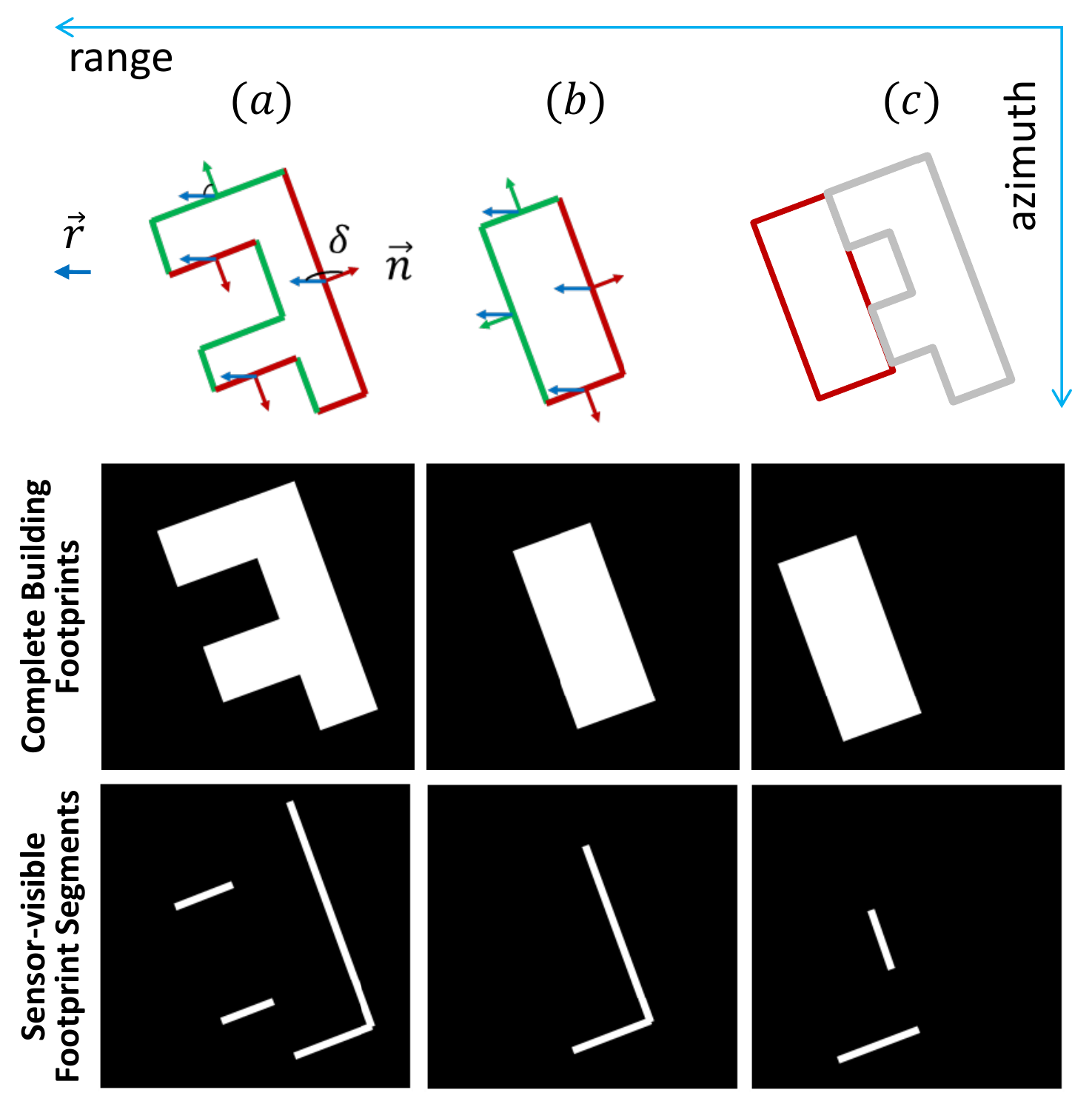}
            \caption{
            Examples of (top) the visibility test of building footprints and (middle and bottom) the two footprint representations. (a) and (b) show footprints of isolated buildings: red edges are sensor-visible, as the angle {$\delta$} between the outward normal vector of an edge {$\protect\overrightarrow{n}$} and the range direction vector {$\protect\overrightarrow{r}$} is in the range of {$(90^{\circ}, 180^{\circ}]$}, while green ones are invisible. (c) shows a case that a footprint is touching another one, hence common edges are sensor-invisible. }
            \label{fig:gisv}
        \end{figure} 
    
    \begin{figure*}[t!]
        \centering
        \includegraphics[width = 0.99\textwidth]{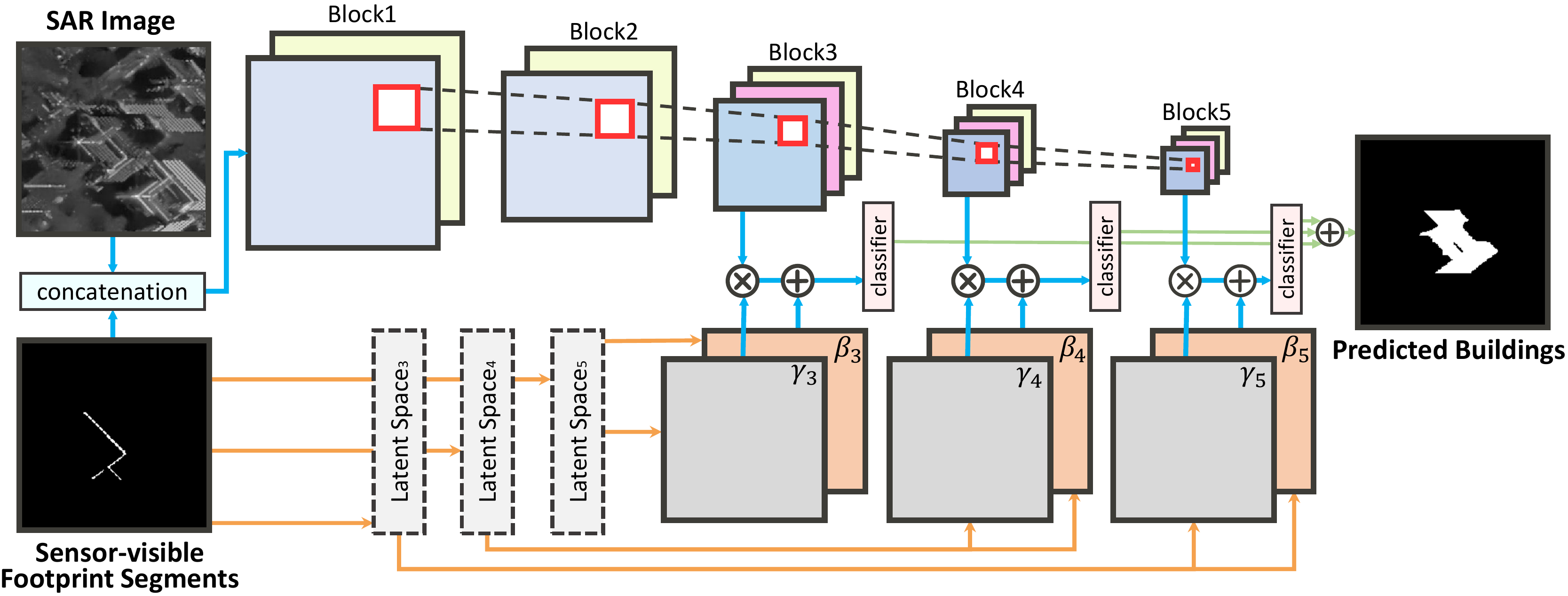}
        \caption{Overview of the CG-Net architecture.}
        \label{fig:netWfcn}
    \end{figure*}       
    
    \subsubsection{Building Distinction}
    
    In this step, we distinguish building points\footnote{Building points refer to points in a point cloud that belong to the building class.} for individual buildings.
    Given one building, 
    its building points are selected from $P_{svs}$ using its footprint. 
    Note that there are two possible inconsistencies between the DEM and GIS data. 
    First, if a building is contained in $P_{svs}$ but not in GIS data, it is not selected from $P_{svs}$. 
    Second, if a building is contained in GIS data but not in $P_{svs}$, i.e., points in the footprint region are not elevated than surrounding ground points, we exclude this building from our dataset. 
    
\subsection{Dataset Generation in the SAR Image Coordinate System}
\label{sec:co_trans}
    \subsubsection{Coordinate Transformation}

    The aforementioned procedures are carried out in the UTM coordinate system, and in our case, building points generated in the previous steps should be projected to the SAR image coordinate system; that is to say, coordinates $(x,y,h)$ need to be transformed to $(range, azimuth)$. 
    Moreover, building footprints are also projected to this coordinate system by using ground height values obtained from the DEM. 
    Generally, the coordinate transformation of the point cloud from UTM coordinate system to the SAR imaging coordinate system includes iterative solving Doppler-Range-Ellipsoid equations that can be implemented with different approaches  \cite{4157311,schwabisch1998fast,toutin2004geometric,roth2004geocoding}. 
    In this work, radar coding was performed using DLR' s Integrated Wide Area Processor (IWAP) \cite{gonzalez2013integrated}.

    \subsubsection{Mask Creation}

    Finally, according to $range-azimuth$ coordinates of building points, we generate building ground truth masks, in which buildings are indicated by 1 and backgrounds are marked as 0. 
    In addition, building footprint masks in the SAR image coordinate system are also created. 
    Notably, in order to find out an effective way of using building footprints, we create two representations, namely complete building footprints and sensor-visible footprint segments. 
    The latter is generated via a visibility test (see Fig. \ref{fig:gisv}). Formally, let $\overrightarrow{n}$ be the outward normal vector of a polygon edge, $\overrightarrow{r}$ be the range direction vector, and $\delta \in [0^{\circ},  180^{\circ}]$ be the angle between $\overrightarrow{n}$ and $\overrightarrow{r}$. A polygon edge is sensor-visible if  $\delta \in (90^{\circ},  180^{\circ}]$, and if a footprint is touching other footprints, common edges are invisible because they do not exist in the real world (e.g., Fig. \ref{fig:gisv}(c)).

    \subsection{Post-processing}
    
    Since the used SAR image and DEM are collected at different times, there might be inconsistencies resulted from urban changes, such as building construction and deconstruction. This leads to inaccurate ground truth data.
    We cope with the problem using intensity values of the given SAR image. 
    In the SAR image, the intensity values are generally larger in building areas than ground areas. 
    Therefore, a threshold is set to be the mode of the intensity values of the SAR image, to exclude buildings of which mean intensity values are smaller than the threshold.

\section{Methodology}\label{sec:CG}

\subsection{Overview}

In this work, our goal is to train a network that takes a SAR image and building footprint as inputs and predicts the building area associated with the footprint in the SAR image. 
Since footprints and visible segments generated from GIS data can provide precise geometry and location information, we resort to exploiting such cues in our task and devise a network module that performs a conditional GIS-aware normalization. By utilizing the CG module, our network, termed as CG-Net, can 
learn feature representations from not only SAR but also GIS data. Specifically, we employ VGG-16 \cite{vgg} as the backbone of CG-Net to learn multi-level features from SAR images. Afterwards, outputs of the last three convolutional blocks are upsampled and fed into the CG module separately. Meanwhile, footprints or visible segments are imported into the CG module as complementary inputs in order to yield final predictions. In what follows, Section \ref{sec:feature_extract} illustrates the procedure of multi-level feature extraction. Section~\ref{sec:cg_module} introduces details of our CG module, and Section~\ref{sec:cg_net} details the configuration of our CG-Net.

    \begin{figure}[t!]
        \centering
        \includegraphics[width = \columnwidth]{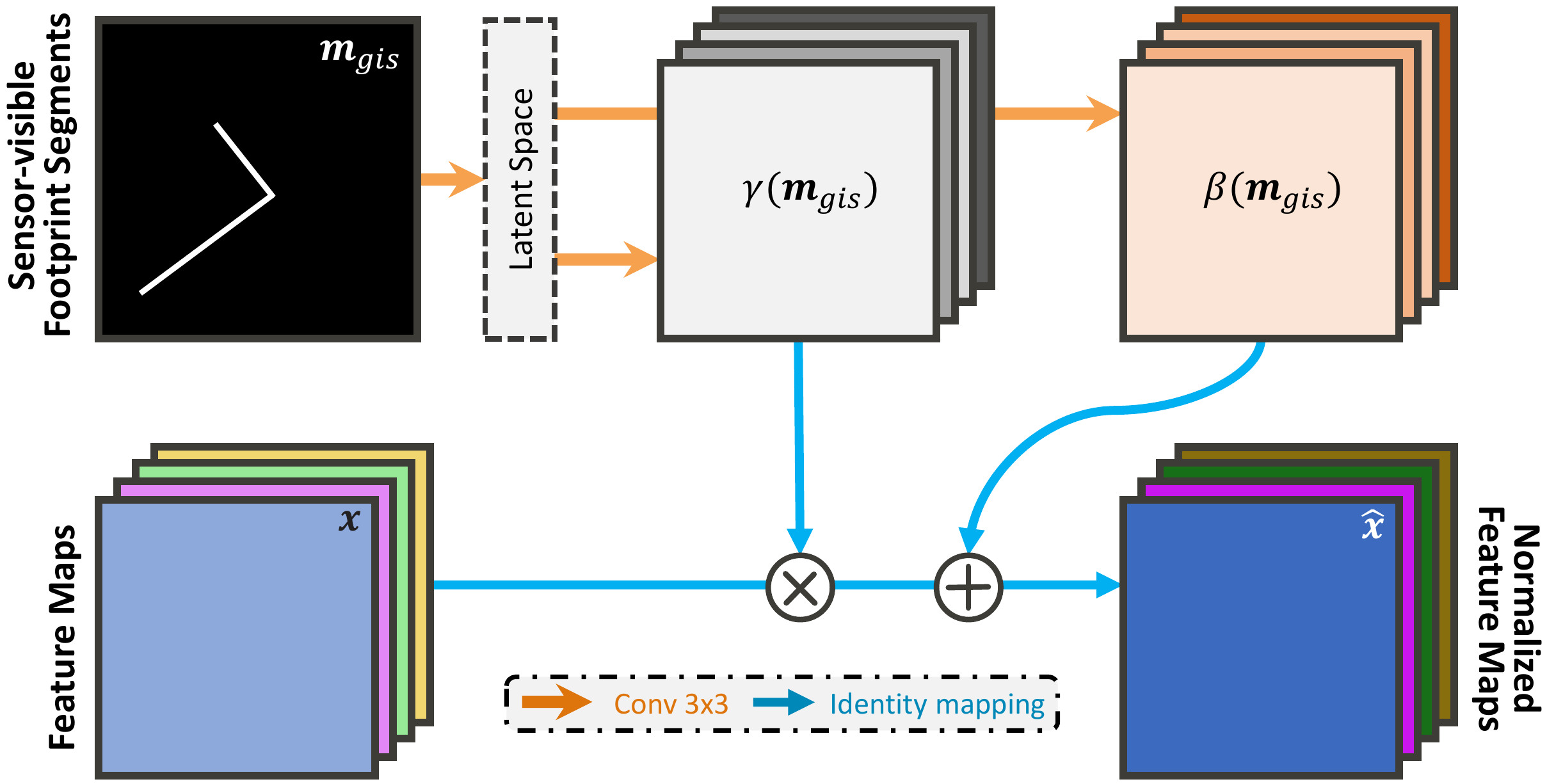}
        \caption{Architecture of the proposed CG module. Here, we take the sensor-visible footprint segments as an example. $\gamma$ and $\beta$ are normalization parameters learned from the sensor-visible footprint segments and used to normalize input feature maps with Eq. (\ref{eq:cg_normalization}).}
        \label{fig:cg}
    \end{figure}
    
\subsection{Multi-level Feature Extraction Module}
\label{sec:feature_extract}

We make use of VGG-16~\cite{vgg} as the backbone of our network to extract features from multiple layers, as these multi-level features help in recognizing buildings with variant scales. 
The backbone consists of five convolutional blocks, and each of them contains two or three convolutional layers. 
The size of their filters is $3 \times 3$. Outputs of all convolutional layers are activated by ReLU~\cite{krizhevsky2012imagenet}, and $2 \times 2$ max-pooling layers with a pooling stride of 2 are interleaved among these blocks. Features learned from deep layers are considered to include high-level semantics, while those from shallow layers are low-level. Therefore, in this task, we utilize features learned from the last three blocks, i.e., \textit{Block3}, \textit{Block4}, and \textit{Block5} (see Fig.~\ref{fig:netWfcn}). Afterwards, the extracted features are fed into the CG module separately.

\subsection{Conditional GIS-aware Normalization Module} 
\label{sec:cg_module}

An intuitive way to make use of GIS data is to simply concatenate them with SAR images and then feed them to a vanilla semantic segmentation network, such as fully convolutional networks (FCN). However, such a method might suffer from the inefficient use of GIS data and leads to unstructured predictions (see the third column in Fig.~\ref{fig:patch}). To address this issue, in this paper, we propose a conditional GIS-aware normalization module to distill the geometry information of individual buildings from GIS data and normalize final predictions with such information.  Formally, let $\bm{m}_{gis}$ be  the mask of the complete building footprint or sensor-visible footprint segments with a spatial size of $W \times H$, and $\bm{x}_b$ denotes feature maps extracted from the $b$-th convolutional block. The width and height of $\bm{x}_b$ are represented as $W'$ and $H'$, respectively. The number of channels is denoted as $C'$. We consider a naive conditional normalization procedure as follows:

\begin{equation}
\label{eq:navive_normalization}
    \hat{\bm{x}}_b = \gamma_b\bm{x}_b+\beta_b,
\end{equation}
where, $\gamma_b$ and $\beta_b$ represent a scale factor and a bias, respectively, and they indicate to what extent $\bm{x}_b$ should be scaled and shifted. The normalized $\bm{x}_b$ is denoted as $\hat{\bm{x}}_b$. A commonly-used measure of $\gamma$ and $\beta$ is to calculate the standard deviation and mean of $\bm{x}_b$. Since $\bm{x}_b$ consists of more than one channel, $\gamma$ and $\beta$ are often computed in a channel-wise manner, and thus, Eq. (\ref{eq:navive_normalization}) can be rewritten as

\begin{equation}
\label{eq:navive_normalization2}
    \hat{\bm{x}}_{b,c} = \gamma_{b,c}(\bm{x}_{b,c})\cdot\bm{x}_{b,c}+\beta_{b,c}(\bm{x}_{b,c}),
\end{equation}
where $c$ denotes the $c$-th channel of $\bm{x}_b$ and ranges from 1 to $C'$. This equation can be easily extended to the batch normalization~\cite{ioffe2015batch} by computing the standard deviation and mean of each $\bm{x}_{b,c}$ in a batch.

In our case, we want to normalize feature representations learned from SAR images, conditioned on GIS data. Our insight is that the GIS data imply coarse localization cues, and their use can guide the network to segment individual buildings accurately. Therefore, we reformulate Eq. (\ref{eq:navive_normalization2}) as follows:

\begin{equation}
\label{eq:cg_normalization}
    \hat{\bm{x}}_{b,c,p,q} = \gamma_{b,c,p,q}(\bm{m}_{gis})\cdot\bm{x}_{b,c,p,q}+\beta_{b,c,p,q}(\bm{m}_{gis}),
\end{equation}
where $\gamma_{b,c,p,q}$ and $\beta_{b,c,p,q}$ indicate the scale factor and bias \textit{learned} specifically for the pixel located at $(p, q)$ in the $c$-th channel of $\bm{x}_b$. As a consequence, normalization parameters $\bm{\gamma}_b$ and $\bm{\beta}_b$ are formatted as matrices with a size of $W' \times H' \times C'$. Such a design enjoys an advantage that normalization parameters are learned in a data-driven manner, and thus these parameters are expected to be more adapted to $\bm{x}_b$. As to the implementation of Eq. (\ref{eq:cg_normalization}), we first project $\bm{m}_{gis}$ onto a latent space through $3 \times 3$ convolutions and then employ two convolutional layers to learn $\gamma_b$ and $\beta_b$ from the encoded $\bm{m}_{gis}$. Subsequently, the element-wise multiplication of $\gamma_b(\bm{m}_{gis})$ and $\bm{x}_b$ is performed, and the output is added to $\beta_b(\bm{m}_{gis})$ pixel by pixel. Fig.~\ref{fig:cg} illustrates the architecture of our CG module.

\subsection{Configuration of CG-Net}
\label{sec:cg_net}

\begin{figure}[!]
        \centering
        \includegraphics[width = \columnwidth]{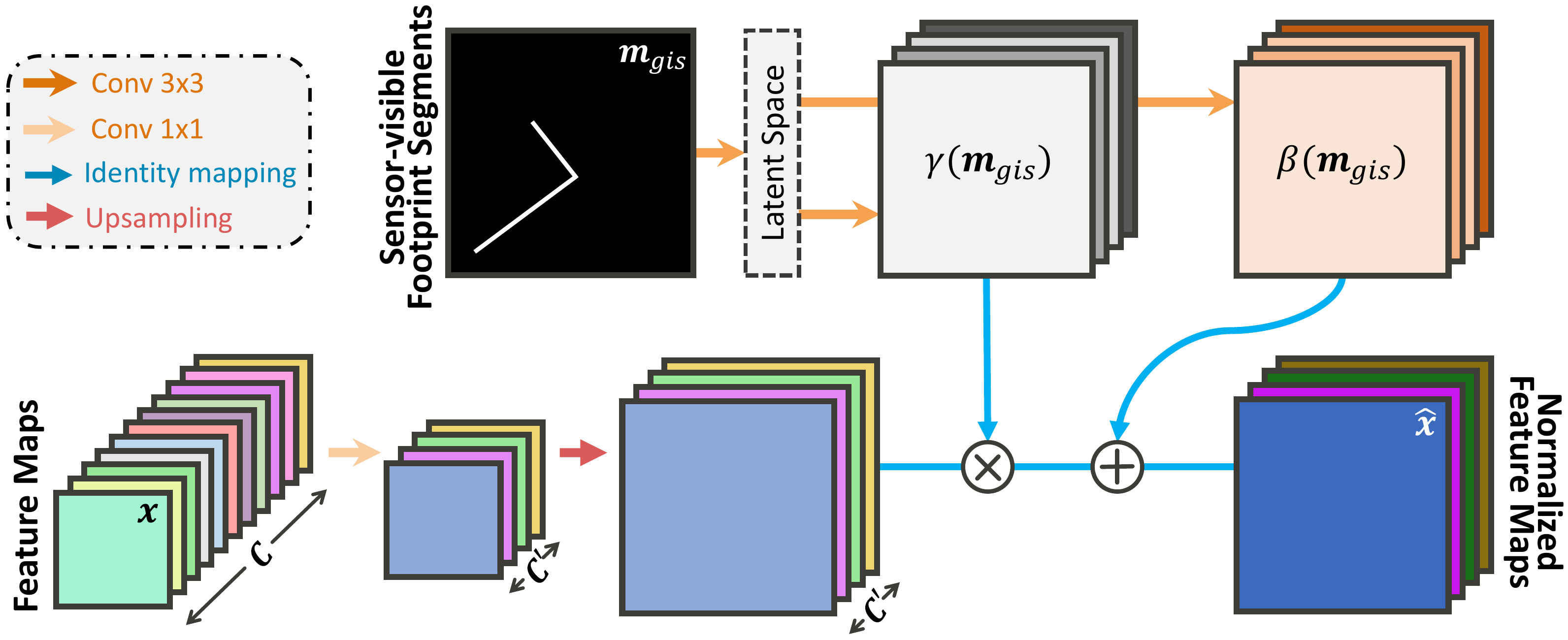}
        \caption{
        Architecture of the final CG module. In advance of performing normalization, the channel of input feature maps is first reduced, and the spatial size is enlarged according to that of sensor-visible footprint segments.}
        \label{fig:cg_new}
    \end{figure}

In order to fully exploit GIS data at multiple scales, we append three CG modules to the last three convolutional blocks of the backbone (see Fig.~\ref{fig:netWfcn}). However, a question is that spatial and channel dimensions of the extracted multi-level features are inconsistent with those of complete building footprints/sensor-visible footprint segments. To address this issue, we upsample these multi-level feature maps to match the spatial resolution of $\bm{m}_{gis}$ via bilinear interpolation. Note that doing so would significantly increase the computation overhead of subsequent operations. Hence we reduce the number of feature channels through $1 \times 1$ convolutions and modify the CG module (see Fig.~\ref{fig:cg_new}) accordingly. Outputs of the CG module are squashed into the number of classes\textcolor{red}{, 2,} and added via an element-wise addition operation to produce final segmentation results. Fig.~\ref{fig:netWfcn} illustrates the architecture of the proposed CG-Net. Furthermore, we note that the proposed CG module is in a plug-and-play fashion and is flexible enough to enhance other semantic segmentation network architectures, e.g., DeepLabv3. For DeepLabv3, since it already fuses features from different layers in its architecture, we simply add our module right before the last layer.

\section{Experiments}\label{sec:test}

\subsection{Data Description}
    
\begin{figure}[!]
            \centering
            \includegraphics[width=\columnwidth]{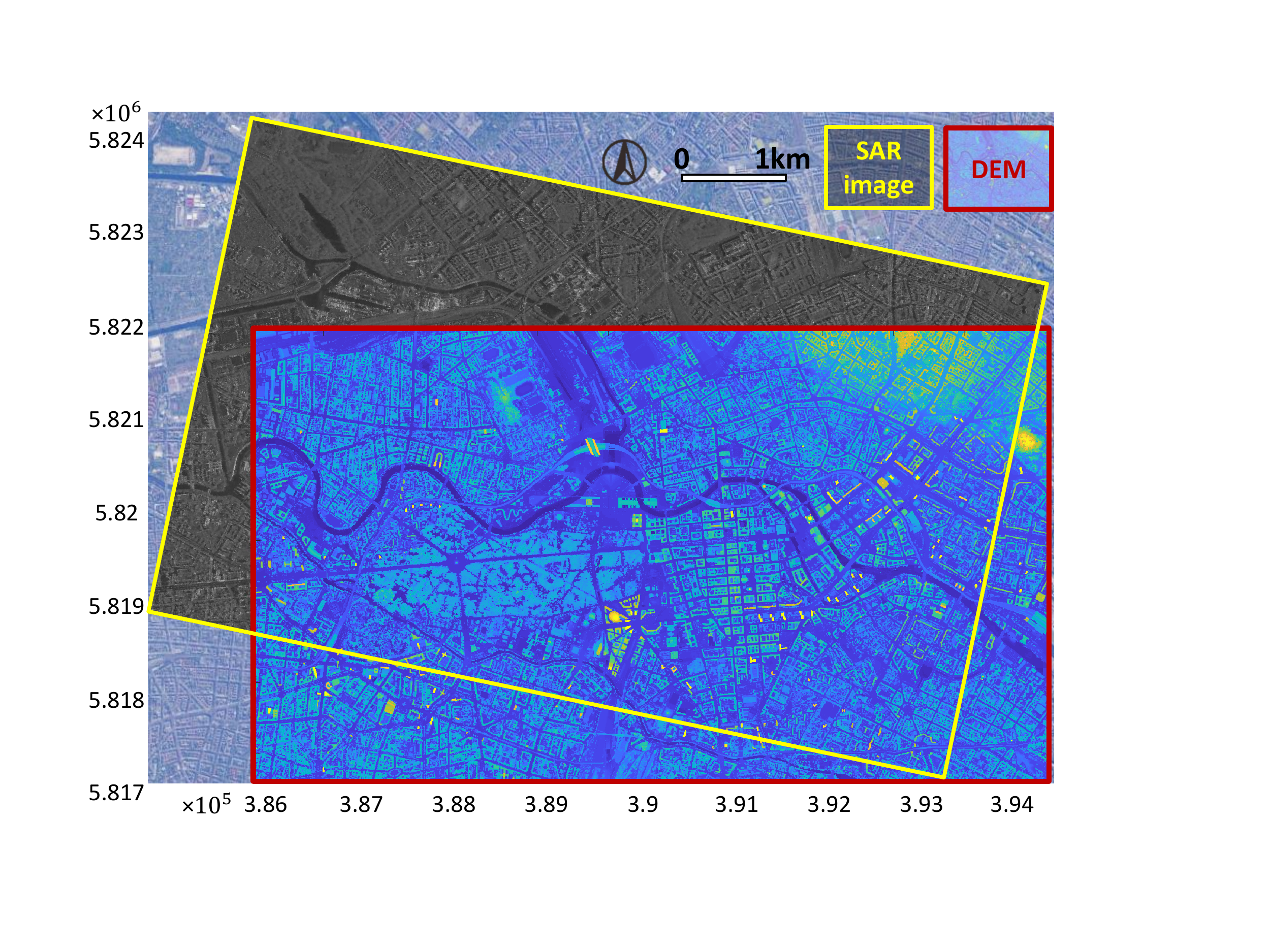}
            \caption{We show our study area in the UTM coordinate system which is the interaction between the SAR image and the DEM.}
            \label{fig:utmarea}
        \end{figure}

In our dataset, a TerraSAR-X image was acquired in the high resolution spotlight mode over Berlin with the pixel spacing\footnote{In SAR images, \textit{pixel spacing} represents the length one pixel corresponds to in the real world, while \textit{resolution} indicates the minimum distance at which the radar can distinguish two close scatters.} of 0.871 m in the azimuth direction and 0.455 m in the slant range direction.  
The incidence angle of this SAR image is 36$^{\circ}$, and the heading angle is 194.34$^{\circ}$. 
To reduce speckle effect, the SAR image was filtered using a nonlocal InSAR algorithm \cite{baierNonlocal}. 
Besides, building footprints in the study area were downloaded from Berlin 3D-Download Portal\footnote{https://www.businesslocationcenter.de/downloadportal/}. 
In order to yield ground truth annotations, we use
a highly accurate DEM that was obtained via the stereo processing of aerial images with a resolution of 7cm/pixel \cite{hirschmuller2008Stereo}. 
Fig. \ref{fig:utmarea} illustrates our study region (the intersection area), the SAR image (yellow rectangle), and DEM (red rectangle). Notably, only data covering the study region are used for generating our dataset.

By using the workflow described in Section \ref{sec:data}, building annotations and footprints are generated. Since we want to explore how GIS data can be effectively used for individual building segmentation, these two versions of footprint masks are produced, namely complete building footprints and sensor-visible footprint segments. Our dataset therefore contains a $5736\times10312$ SAR image, two versions of footprint masks, and ground truths of individual buildings.

    \newlength{\tempdima}
\newcommand{\rowname}[1]
{\rotatebox{90}{\makebox[\tempdima][c]{\textbf{#1}}}}

\renewcommand{\thesubfigure}{\alph{subfigure}}
\newcommand{\mycaption}[1]
{\refstepcounter{subfigure}\textbf{(\thesubfigure) }{\ignorespaces #1}}

\begin{figure*}[!]
    \centering
    \settoheight{\tempdima}{\includegraphics[width=.15\textwidth]{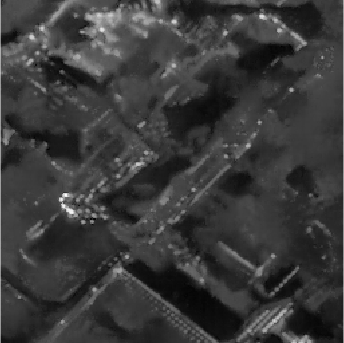}}%
    \begin{tabularx}{0.99\linewidth}{c@{\hskip2pt}c@{\hskip2pt}c@{\hskip2pt}c@{\hskip2pt}c@{\hskip2pt}c@{\hskip2pt}c@{\hskip2pt}c@{\hskip2pt}c@{\hskip2pt}}
    \\\vspace{-0.35cm}
    \textbf{\footnotesize{a}} &\textbf{\footnotesize{b}}& \textbf{\footnotesize{c}} & \textbf{\footnotesize{d}} & \textbf{\footnotesize{e}} & \textbf{\footnotesize{f}} & \textbf{\footnotesize{g}}
    \\\vspace{-0.4cm}
    \rowname{\hspace{-0.3cm}\footnotesize{SAR image}}
    \subfloat{\reflectbox{\includegraphics[width=.13\textwidth]{figs/subfigs/00554_01.png}}}&
    \subfloat{\reflectbox{\includegraphics[width=.13\textwidth]{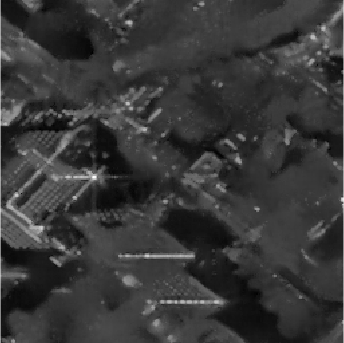}}}& 
    \subfloat{\reflectbox{\includegraphics[width=.13\textwidth]{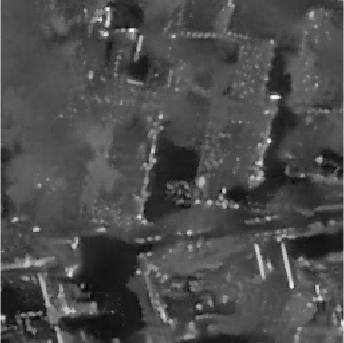}}}&
    \subfloat{\reflectbox{\includegraphics[width=.13\textwidth]{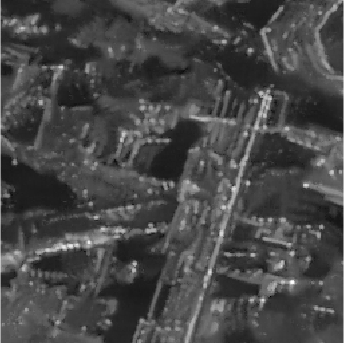}}}&
    \subfloat{\reflectbox{\includegraphics[width=.13\textwidth]{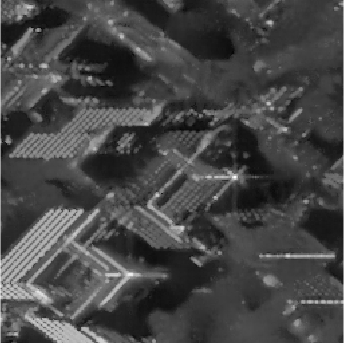}}}&
    \subfloat{\reflectbox{\includegraphics[width=.13\textwidth]{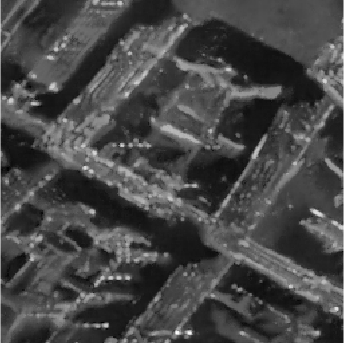}}}&
    \subfloat{\reflectbox{\includegraphics[width=.13\textwidth]{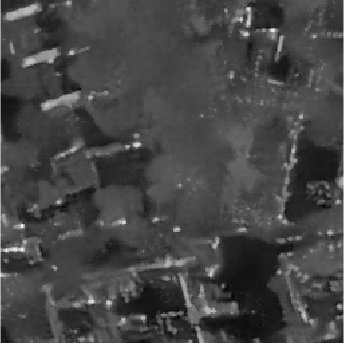}}}&
    \\\vspace{-0.4cm}
    \rowname{\hspace{-0.3cm}\footnotesize{SFS}}             \subfloat{\reflectbox{\includegraphics[width=.13\textwidth]{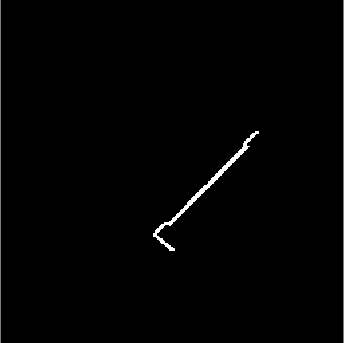}}}&
    \subfloat{\reflectbox{\includegraphics[width=.13\textwidth]{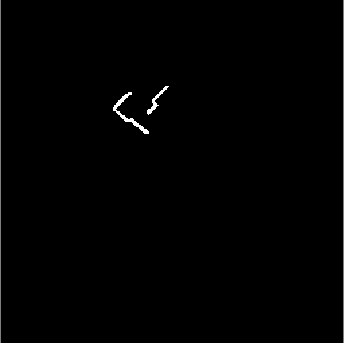}}}&
    \subfloat{\reflectbox{\includegraphics[width=.13\textwidth]{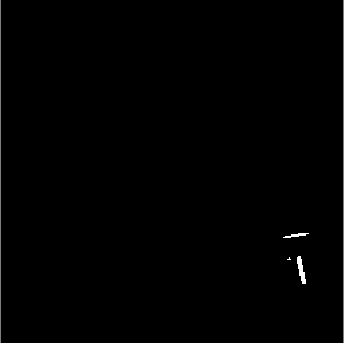}}}&
    \subfloat{\reflectbox{\includegraphics[width=.13\textwidth]{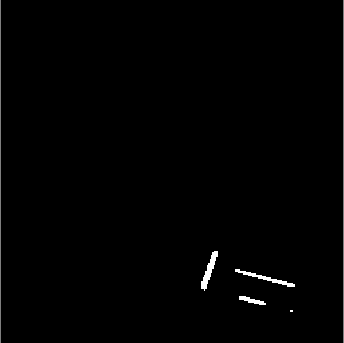}}}&
    \subfloat{\reflectbox{\includegraphics[width=.13\textwidth]{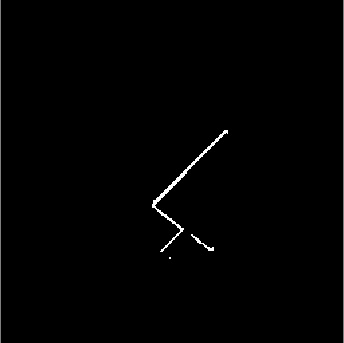}}}&
    \subfloat{\reflectbox{\includegraphics[width=.13\textwidth]{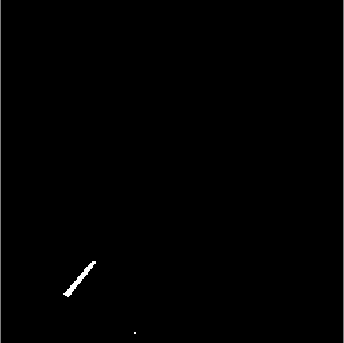}}}&
    \subfloat{\reflectbox{\includegraphics[width=.13\textwidth]{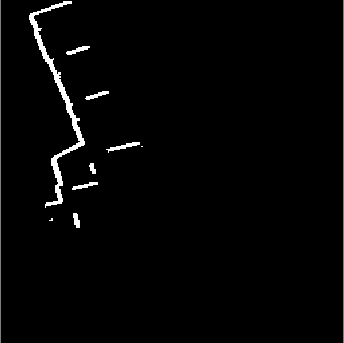}}}&
    \\\vspace{-0.4cm}
    \rowname{\hspace{-0.3cm}\footnotesize{FCN}}
    \subfloat{\reflectbox{\includegraphics[width=.13\textwidth]{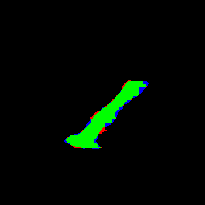}}}&
    \subfloat{\reflectbox{\includegraphics[width=.13\textwidth]{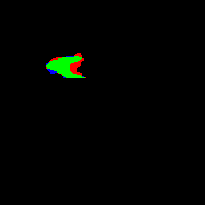}}}&
    \subfloat{\reflectbox{\includegraphics[width=.13\textwidth]{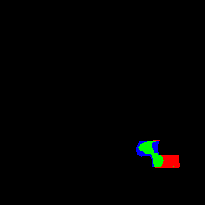}}}&
    \subfloat{\reflectbox{\includegraphics[width=.13\textwidth]{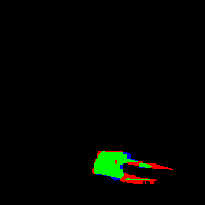}}}&
    \subfloat{\reflectbox{\includegraphics[width=.13\textwidth]{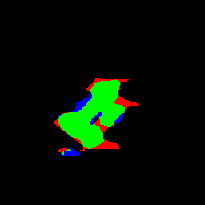}}}&
    \subfloat{\reflectbox{\includegraphics[width=.13\textwidth]{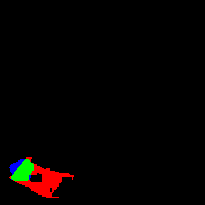}}}&
    \subfloat{\reflectbox{\includegraphics[width=.13\textwidth]{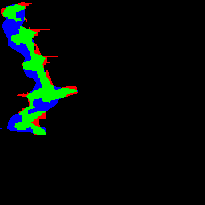}}}&
    \\\vspace{-0.4cm}
    \rowname{\hspace{-0.3cm}\footnotesize{FCN-CG}} 
    \subfloat{\reflectbox{\includegraphics[width=.13\textwidth]{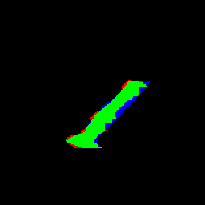}}}&
    \subfloat{\reflectbox{\includegraphics[width=.13\textwidth]{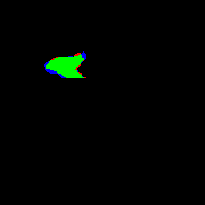}}}&
    \subfloat{\reflectbox{\includegraphics[width=.13\textwidth]{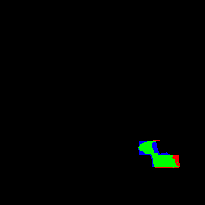}}}&
    \subfloat{\reflectbox{\includegraphics[width=.13\textwidth]{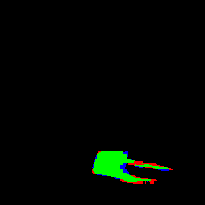}}}&
    \subfloat{\reflectbox{\includegraphics[width=.13\textwidth]{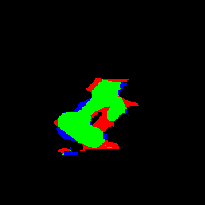}}}&
    \subfloat{\reflectbox{\includegraphics[width=.13\textwidth]{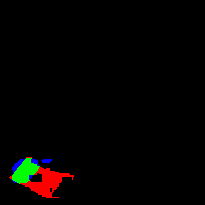}}}&
    \subfloat{\reflectbox{\includegraphics[width=.13\textwidth]{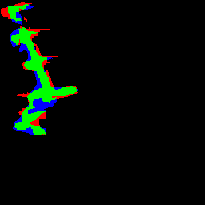}}}&
    \\\vspace{-0.4cm}
    \rowname{\hspace{-0.3cm}\footnotesize{DeepLabv3}} 
    \subfloat{\reflectbox{\includegraphics[width=.13\textwidth]{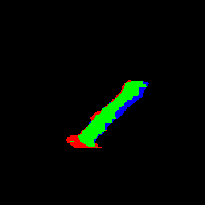}}}&
    \subfloat{\reflectbox{\includegraphics[width=.13\textwidth]{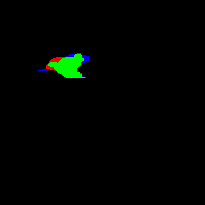}}}&
    \subfloat{\reflectbox{\includegraphics[width=.13\textwidth]{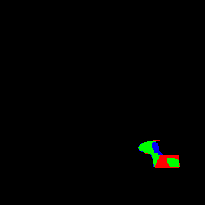}}}&
    \subfloat{\reflectbox{\includegraphics[width=.13\textwidth]{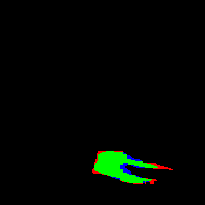}}}&
    \subfloat{\reflectbox{\includegraphics[width=.13\textwidth]{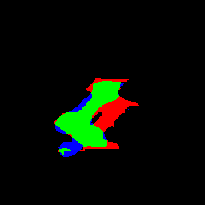}}}&
    \subfloat{\reflectbox{\includegraphics[width=.13\textwidth]{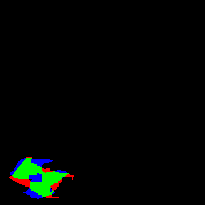}}}&
    \subfloat{\reflectbox{\includegraphics[width=.13\textwidth]{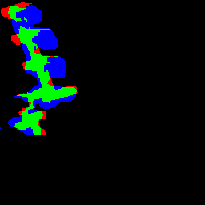}}}&
    \\\vspace{-0.4cm}
    \rowname{\hspace{-0.3cm}\footnotesize{DeepLabv3-CG}} 
    \subfloat{\reflectbox{\includegraphics[width=.13\textwidth]{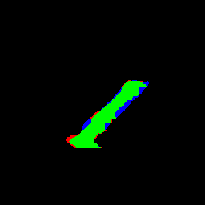}}}&
    \subfloat{\reflectbox{\includegraphics[width=.13\textwidth]{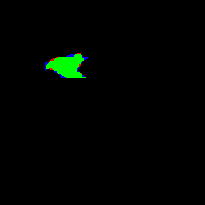}}}&
    \subfloat{\reflectbox{\includegraphics[width=.13\textwidth]{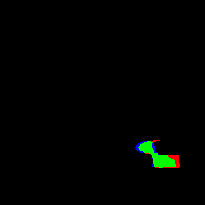}}}&
    \subfloat{\reflectbox{\includegraphics[width=.13\textwidth]{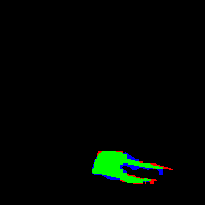}}}&
    \subfloat{\reflectbox{\includegraphics[width=.13\textwidth]{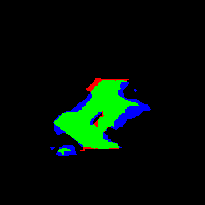}}}&
    \subfloat{\reflectbox{\includegraphics[width=.13\textwidth]{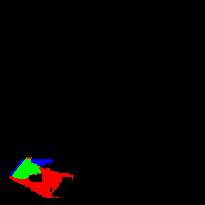}}}&
    \subfloat{\reflectbox{\includegraphics[width=.13\textwidth]{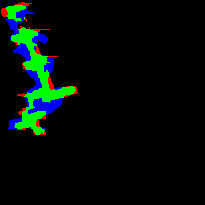}}}&\\
    & & & & & & &\\
    \end{tabularx}
    \caption{Examples of segmentation results using sensor-visible footprint segments (abbreviated as SFS). Pixel-based true positives, false positives, and false negatives are marked in green, red, and blue, respectively.}
    \label{fig:subgisV}
\end{figure*}
    \begin{figure*}[!]
    \centering
    \settoheight{\tempdima}{\includegraphics[width=.15\textwidth]{figs/subfigs/00554_01.png}}%
    \begin{tabularx}{0.99\linewidth}{c@{\hskip2pt}c@{\hskip2pt}c@{\hskip2pt}c@{\hskip2pt}c@{\hskip2pt}c@{\hskip2pt}c@{\hskip2pt}c@{\hskip2pt}c@{\hskip2pt}}
    \\\vspace{-0.35cm}
    \textbf{\footnotesize{a}} &\textbf{\footnotesize{b}}& \textbf{\footnotesize{c}} & \textbf{\footnotesize{d}} & \textbf{\footnotesize{e}} & \textbf{\footnotesize{f}} & \textbf{\footnotesize{g}}
    \\\vspace{-0.4cm}
    \rowname{\hspace{-0.3cm}\footnotesize{SAR image}}
    \subfloat{\reflectbox{\includegraphics[width=.13\textwidth]{figs/subfigs/00554_01.png}}}&
    \subfloat{\reflectbox{\includegraphics[width=.13\textwidth]{figs/subfigs/00524_01.png}}}& 
    \subfloat{\reflectbox{\includegraphics[width=.13\textwidth]{figs/subfigs/10589_01.png}}}&
    \subfloat{\reflectbox{\includegraphics[width=.13\textwidth]{figs/subfigs/02539_01.png}}}&
    \subfloat{\reflectbox{\includegraphics[width=.13\textwidth]{figs/subfigs/00515_01.png}}}&
    \subfloat{\reflectbox{\includegraphics[width=.13\textwidth]{figs/subfigs/00313_01.png}}}&
    \subfloat{\reflectbox{\includegraphics[width=.13\textwidth]{figs/subfigs/10586_01.png}}}&
    \\\vspace{-0.4cm}
    \rowname{\hspace{-0.3cm}\footnotesize{CBF}}
    \subfloat{\reflectbox{\includegraphics[width=.13\textwidth]{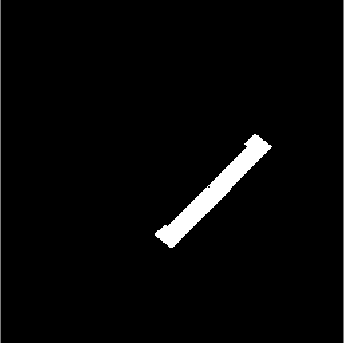}}}&
    \subfloat{\reflectbox{\includegraphics[width=.13\textwidth]{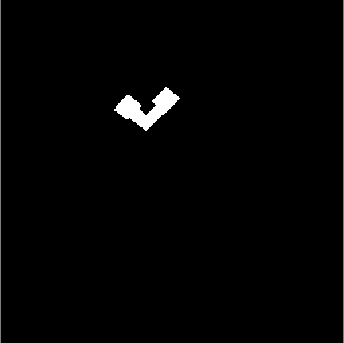}}}&
    \subfloat{\reflectbox{\includegraphics[width=.13\textwidth]{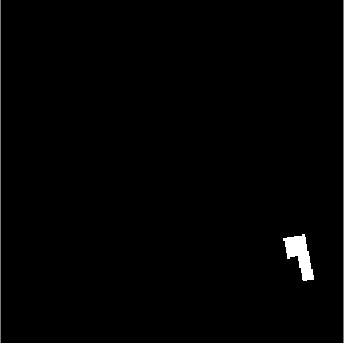}}}&
    \subfloat{\reflectbox{\includegraphics[width=.13\textwidth]{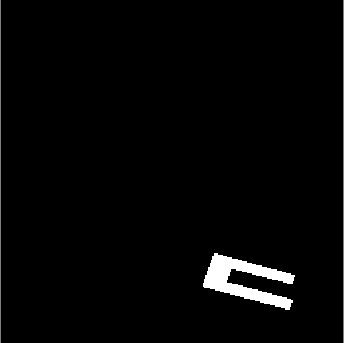}}}&
    \subfloat{\reflectbox{\includegraphics[width=.13\textwidth]{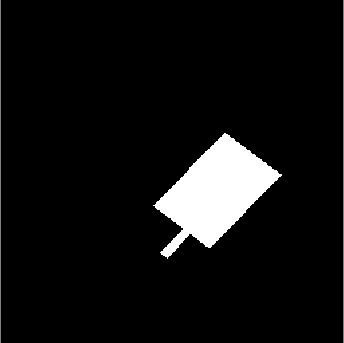}}}&
    \subfloat{\reflectbox{\includegraphics[width=.13\textwidth]{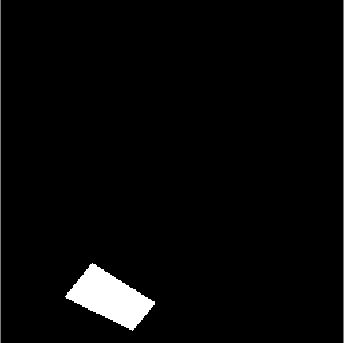}}}&
    \subfloat{\reflectbox{\includegraphics[width=.13\textwidth]{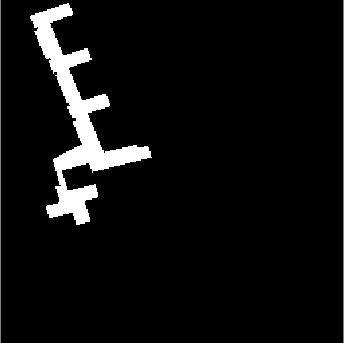}}}&
    \\\vspace{-0.4cm}
    \rowname{\hspace{-0.3cm}\footnotesize{FCN}}
    \subfloat{\reflectbox{\includegraphics[width=.13\textwidth]{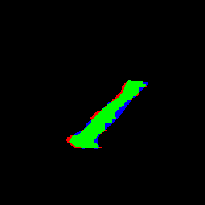}}}&
    \subfloat{\reflectbox{\includegraphics[width=.13\textwidth]{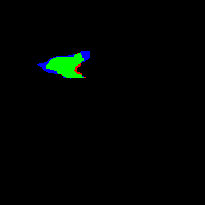}}}&
    \subfloat{\reflectbox{\includegraphics[width=.13\textwidth]{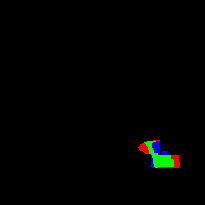}}}&
    \subfloat{\reflectbox{\includegraphics[width=.13\textwidth]{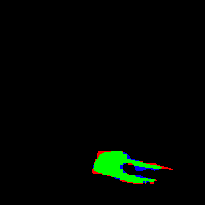}}}&
    \subfloat{\reflectbox{\includegraphics[width=.13\textwidth]{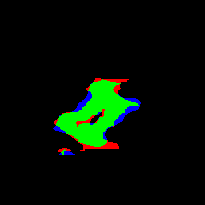}}}&
    \subfloat{\reflectbox{\includegraphics[width=.13\textwidth]{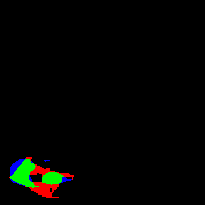}}}&
    \subfloat{\reflectbox{\includegraphics[width=.13\textwidth]{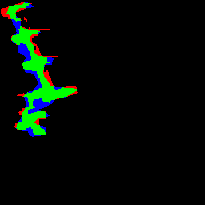}}}&
    \\\vspace{-0.4cm}
    \rowname{\hspace{-0.3cm}\footnotesize{FCN-CG}} 
    \subfloat{\reflectbox{\includegraphics[width=.13\textwidth]{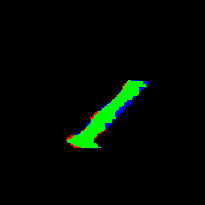}}}&
    \subfloat{\reflectbox{\includegraphics[width=.13\textwidth]{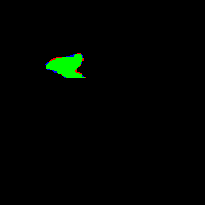}}}&
    \subfloat{\reflectbox{\includegraphics[width=.13\textwidth]{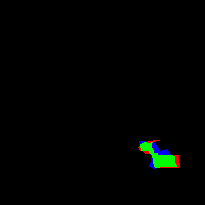}}}&
    \subfloat{\reflectbox{\includegraphics[width=.13\textwidth]{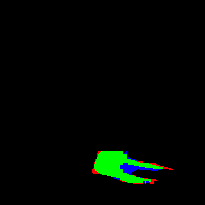}}}&
    \subfloat{\reflectbox{\includegraphics[width=.13\textwidth]{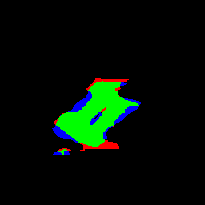}}}&
    \subfloat{\reflectbox{\includegraphics[width=.13\textwidth]{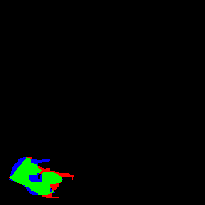}}}&
    \subfloat{\reflectbox{\includegraphics[width=.13\textwidth]{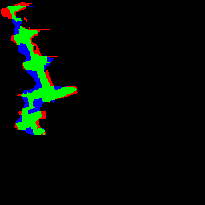}}}&
    \\\vspace{-0.4cm}
    \rowname{\hspace{-0.3cm}\footnotesize{DeepLabv3}}
    \subfloat{\reflectbox{\includegraphics[width=.13\textwidth]{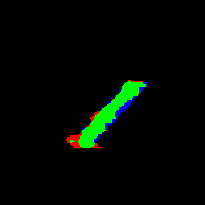}}}&
    \subfloat{\reflectbox{\includegraphics[width=.13\textwidth]{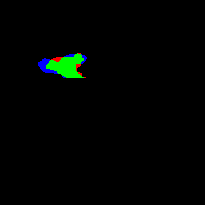}}}&
    \subfloat{\reflectbox{\includegraphics[width=.13\textwidth]{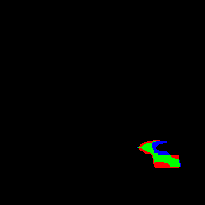}}}&
    \subfloat{\reflectbox{\includegraphics[width=.13\textwidth]{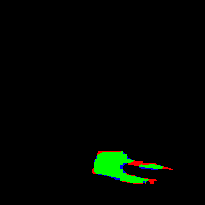}}}&
    \subfloat{\reflectbox{\includegraphics[width=.13\textwidth]{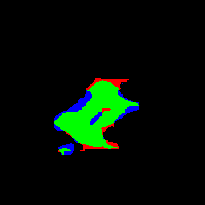}}}&
    \subfloat{\reflectbox{\includegraphics[width=.13\textwidth]{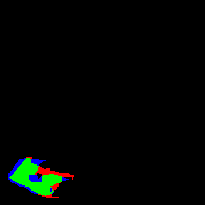}}}&
    \subfloat{\reflectbox{\includegraphics[width=.13\textwidth]{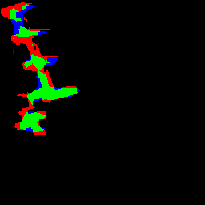}}}&
    \\\vspace{-0.4cm}
    \rowname{\hspace{-0.3cm}\footnotesize{DeepLabv3-CG}} 
    \subfloat{\reflectbox{\includegraphics[width=.13\textwidth]{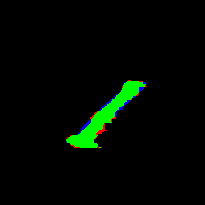}}}&
    \subfloat{\reflectbox{\includegraphics[width=.13\textwidth]{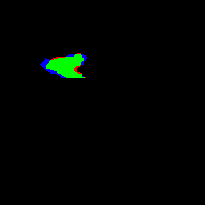}}}&
    \subfloat{\reflectbox{\includegraphics[width=.13\textwidth]{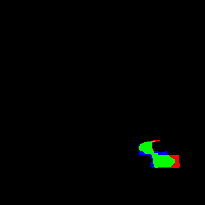}}}&
    \subfloat{\reflectbox{\includegraphics[width=.13\textwidth]{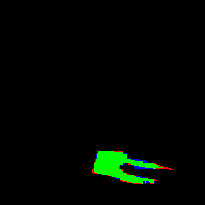}}}&
    \subfloat{\reflectbox{\includegraphics[width=.13\textwidth]{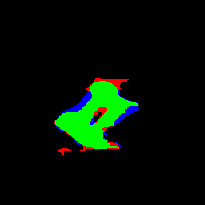}}}&
    \subfloat{\reflectbox{\includegraphics[width=.13\textwidth]{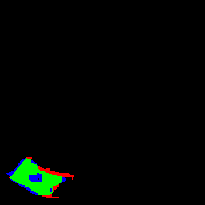}}}&
    \subfloat{\reflectbox{\includegraphics[width=.13\textwidth]{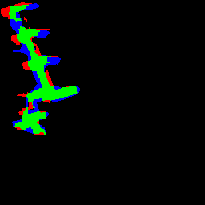}}}&\\
    & & & & & & &\\
    \end{tabularx}
    \caption{Examples of segmentation results using complete building footprints (abbreviated as CBF). Pixel-based true positives, false positives, and false negatives are marked in green, red, and blue, respectively.}
    \label{fig:subgisF}
\end{figure*}
    \begin{figure*}[!]
    \centering
    \begin{tabularx}{0.8\linewidth}{@{}c@{\hskip1pt}c@{\hskip1pt}c@{\hskip1pt}c@{\hskip1pt}c@{\hskip1pt}c@{\hskip1pt}c@{\hskip1pt}}
    \\\vspace{-0.4cm}
    \hspace{0.5cm}\textbf{\footnotesize{SAR image}} & \textbf{\footnotesize{SFS} {\scriptsize(overlaid on GT)}} & \textbf{\footnotesize{FCN}} & \textbf{\footnotesize{FCN-CG}} & \textbf{\footnotesize{DeepLabv3}} & \textbf{\footnotesize{DeepLabv3-CG}}
    \\\vspace{-0.5cm}
    \rowname{\hspace{-0.4cm}\small{patch 1}}
    \subfloat{\reflectbox{\includegraphics[width=.13\textwidth]{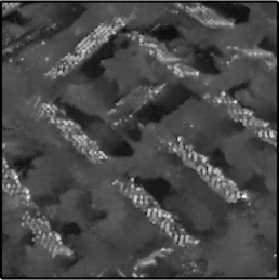}}}&   \subfloat{\reflectbox{\includegraphics[width=.13\textwidth]{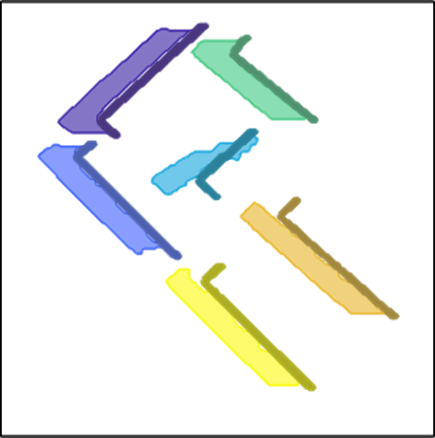}}}&
    \subfloat{\reflectbox{\includegraphics[width=.13\textwidth]{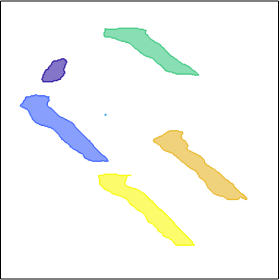}}}& 
    \subfloat{\reflectbox{\includegraphics[width=.13\textwidth]{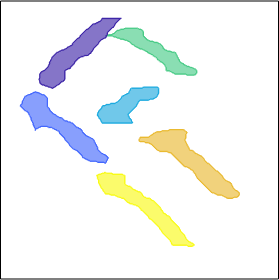}}}& 
    \subfloat{\reflectbox{\includegraphics[width=.13\textwidth]{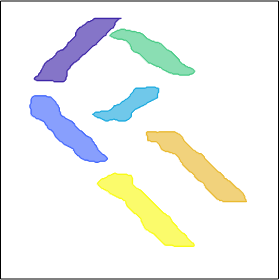}}}& 
    \subfloat{\reflectbox{\includegraphics[width=.13\textwidth]{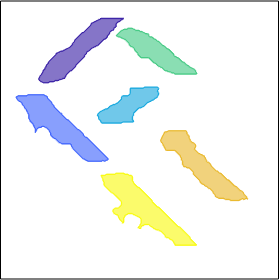}}}
    \\
    & & & & &  &\\\vspace{-0.4cm}
    \hspace{0.5cm}\textbf{\footnotesize{GT}} & \textbf{\footnotesize{CBF} {\scriptsize(overlaid on GT)}} & \textbf{\footnotesize{FCN}} & \textbf{\footnotesize{FCN-CG}} & \textbf{\footnotesize{DeepLabv3}} & \textbf{\footnotesize{DeepLabv3-CG}}
    \\
    \hspace{0.45cm}
    \subfloat{\reflectbox{\includegraphics[width=.13\textwidth]{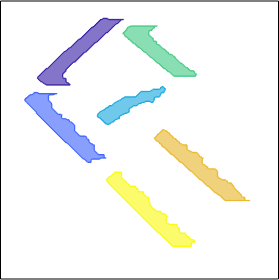}}}& 
    \subfloat{\reflectbox{\includegraphics[width=.13\textwidth]{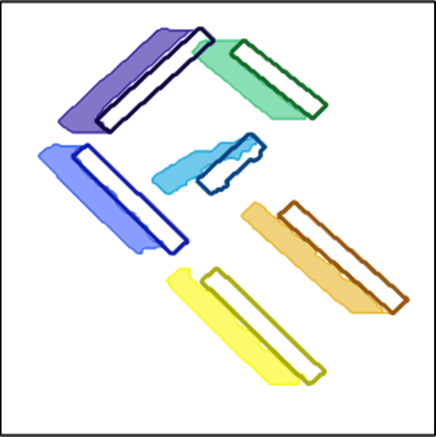}}}&
    \subfloat{\reflectbox{\includegraphics[width=.13\textwidth]{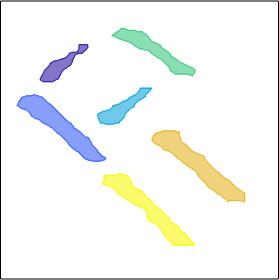}}}& 
    \subfloat{\reflectbox{\includegraphics[width=.13\textwidth]{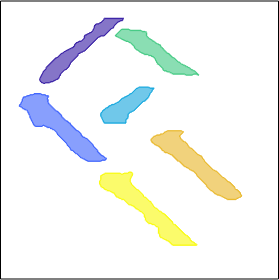}}}& 
    \subfloat{\reflectbox{\includegraphics[width=.13\textwidth]{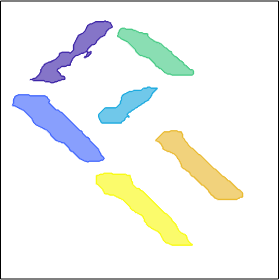}}}& 
    \subfloat{\reflectbox{\includegraphics[width=.13\textwidth]{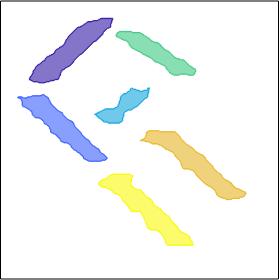}}} 
    \\
     \vspace{-0.4cm}
    \hspace{0.5cm}\textbf{\footnotesize{SAR image}} & \textbf{\footnotesize{SFS} {\scriptsize(overlaid on GT)}} & \textbf{\footnotesize{FCN}} & \textbf{\footnotesize{FCN-CG}} & \textbf{\footnotesize{DeepLabv3}} & \textbf{\footnotesize{DeepLabv3-CG}}
    \\\vspace{-0.4cm}
    \rowname{\hspace{-0.4cm}\small{patch 2}} 
    \subfloat{\reflectbox{\includegraphics[width=.13\textwidth]{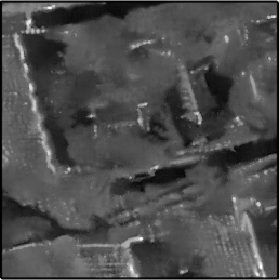}}}&
    \subfloat{\reflectbox{\includegraphics[width=.13\textwidth]{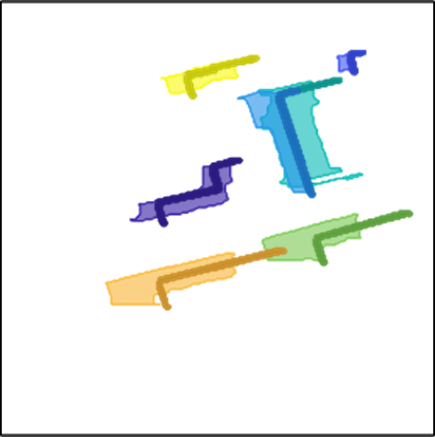}}}&
    \subfloat{\reflectbox{\includegraphics[width=.13\textwidth]{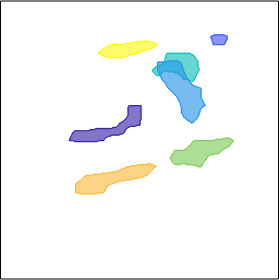}}}& 
    \subfloat{\reflectbox{\includegraphics[width=.13\textwidth]{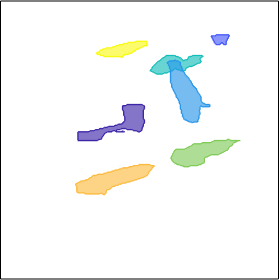}}}& 
    \subfloat{\reflectbox{\includegraphics[width=.13\textwidth]{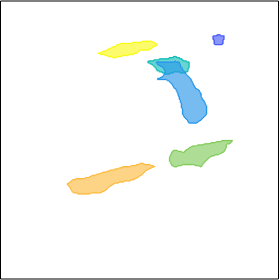}}}& 
    \subfloat{\reflectbox{\includegraphics[width=.13\textwidth]{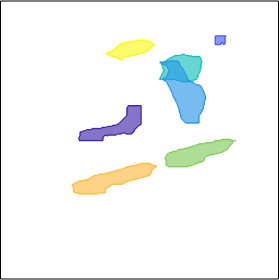}}}
    \\
     & & & & &  &\\\vspace{-0.4cm}
    \hspace{0.5cm}\textbf{\footnotesize{GT}} & \textbf{\footnotesize{CBF} {\scriptsize(overlaid on GT)}} & \textbf{\footnotesize{FCN}} & \textbf{\footnotesize{FCN-CG}} & \textbf{\footnotesize{DeepLabv3}} & \textbf{\footnotesize{DeepLabv3-CG}}
    \\
    \hspace{0.45cm}
    \subfloat{\reflectbox{\includegraphics[width=.13\textwidth]{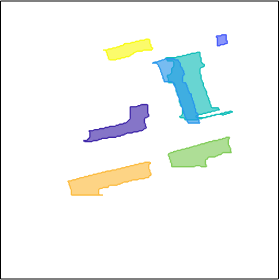}}}&
    \subfloat{\reflectbox{\includegraphics[width=.13\textwidth]{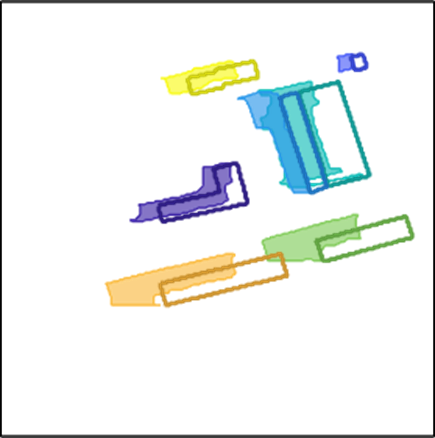}}}&
    \subfloat{\reflectbox{\includegraphics[width=.13\textwidth]{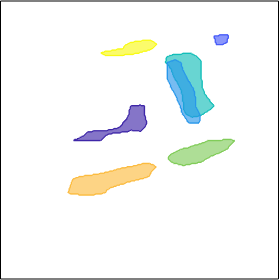}}}& 
    \subfloat{\reflectbox{\includegraphics[width=.13\textwidth]{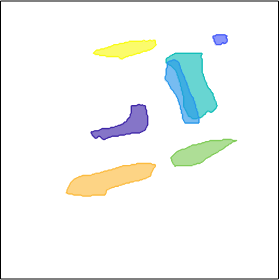}}}& 
    \subfloat{\reflectbox{\includegraphics[width=.13\textwidth]{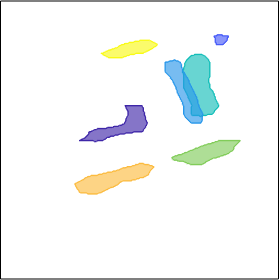}}}& 
    \subfloat{\reflectbox{\includegraphics[width=.13\textwidth]{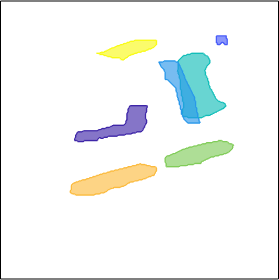}}}
    \\
    \end{tabularx}
    \caption{Examples of segmentation results from different models on two patches, using complete building footprints (abbreviated as CBF) and sensor-visible footprint segments (abbreviated as SFS). CBF and SFS are overlaid on the ground truth (GT) to visualize the difference between building footprints and buildings. Different buildings are plotted in different colors (50\% transparency). }
    \label{fig:patch}
\end{figure*}

\subsection{Training Details} \label{sec:train_details}

In order to train an effective and robust segmentation network, 
we crop the SAR image into patches of $256 \times 256$ pixels with a stride of $150$ pixels. 
Note that patches including incomplete footprints or ground truth annotations are discarded.
Consequently, 30056 buildings are remaining, and each of them has three patches: 
a SAR image patch, a footprint patch, and a ground truth mask. 
Among all buildings, 19434 of them are utilized to train networks, and the others are test samples. 
Note that training and test regions do not overlap. 
The network takes one SAR patch and the corresponding GIS patch for one building as inputs. After predicting masks of all buildings, overlapping areas are  obtained by overlaying all masks.  

During the training phase, components of the proposed CG-Net are initialized with different strategies. Specifically, the multi-level feature extraction module is initialized with weights pre-trained on ImageNet~\cite{imagenet_cvpr09}, and all convolutional layers in the CG modules are initialized with a Glorot uniform initializer. 
The network is implemented on TensorFlow and trained on one NVIDIA Tesla P100 16GB GPU for 155k 
iterations. During the training procedure, all weights are updated through back-propagation, and we select Netrov Adam~\cite{nadam2} as the optimizer. Parameters of this optimizer are set as recommended: $\epsilon=1\mathrm{e}{-08}$, $\beta_1=0.9$, and $\beta_2=0.999$. The loss is defined as binary cross-entropy, as only two classes are considered in our dataset, i.e., building segments and background. We initialize the learning rate as $2e-3$ and reduce it by a factor of $\sqrt{10}$ once the loss stops to decrease for two epochs. Moreover, we utilize a small batch size of 5 in our experiments.

\subsection{Quantitative Evaluation}

To evaluate the performance of networks, we calculate the F1 score as follows:
\begin{equation}
    F1 = 2\cdot\frac{P \cdot R}{P + R}, 
    P = \frac{tp}{tp + fp}, 
    R = \frac{tp}{tp + fn},
\end{equation}
where $P$ and $R$ denote the precision and recall, respectively. 
In addition, the intersection over union (IoU) and overall accuracy (OA) are also calculated for a comprehensive comparison:
\begin{equation}
    IoU = \frac{tp}{tp+fp+fn},
    OA = \frac{tp+tn}{tp+tn+fp+fn}. 
\end{equation}
$tp$, $fp$, $tn$, $fn$ represent pixel-based true positives, false positives, true negatives, and false negatives for buildings, respectively.

In our experiments, we compare four models: FCN, FCN-CG, DeepLabv3, and DeepLabv3-CG. 
It is worth mentioning that FCN and DeepLabv3 are regarded as baselines, and their inputs are concatenations of SAR patches and their corresponding footprint patches. Both FCN-CG and DeepLabv3-CG are our proposed networks with different backbones.

Table \ref{tab:gisV} reports numerical results of different models on our dataset, where sensor-visible footprint segments are used. 
Comparison of these results corroborates that the proposed CG module can improve the performance of individual building segmentation.
Specifically, compared to FCN and DeepLabv3, FCN-CG and DeepLabv3-CG achieve improvements of 0.75\% and 2.17\% in the precision, respectively. Besides, increments of 1.23\% and 1.65\% in the mean F1 score and IoU can be observed by comparing FCN-CG and FCN, while improvements of 0.97\% and 1.14\% in the same metrics are achieved by introducing the CG module to DeepLabv3. 

Table \ref{tab:gisF} presents results of variant models using complete building footprints. 
We can see that the results are consistent with those using sensor-visible footprint segments. 
For example, with the CG module, the precision improves 1.95\% and 3.94\% with the backbone, FCN and DeepLabv3, and the IoU increases 1.50\% and 2.16\%. 
To summarize, improvements achieved by FCN-CG and DeepLabv3-CG demonstrate the effectiveness of the proposed CG module, 
and DeepLabv3-CG can achieve the best performance in all four metrics on our dataset.
Moreover, we note that all models achieve relatively high OAs, and even the worst model can achieve an OA of 83.40\%. This is because OA is computed by considering all pixels, while non-building pixels, which are easily recognized, account for a large proportion.

        \begin{table}[h]
            \centering
            \caption{
            Numerical results using sensor-visible footprint segments. The highest values of different metrics are highlighted in \textbf{bold}.}\label{tab:gisV}
            \begin{tabular}{@{\hskip8pt}c@{\hskip8pt}c@{\hskip8pt}c@{\hskip8pt}c@{\hskip8pt}c@{\hskip8pt}c@{\hskip8pt}c@{\hskip8pt}c@{\hskip8pt}c@{\hskip8pt}c}
                \hline
                Model Name            & P    & F1 score   & IoU      & OA     \\
                \hline
                FCN                   & 0.6478          & 0.6808  & 0.5138      & 0.8340 \\ 
                FCN-CG                & 0.6553          & 0.6931  & 0.5303      & 0.9926 \\
                DeepLabv3             & 0.6635          & 0.6971  & 0.5351      & 0.9927 \\
                DeepLabv3-CG          & \textbf{0.6852}        & \textbf{0.7068}  & \textbf{0.5465}   & \textbf{0.9928} \\
                \hline
            \end{tabular}
        \end{table}

        \begin{table}[!]
            \centering
            \caption{Numerical results using complete building footprints.  The highest values of different metrics are highlighted in \textbf{bold}.}\label{tab:gisF}
            \begin{tabular}{@{\hskip8pt}c@{\hskip8pt}c@{\hskip8pt}c@{\hskip8pt}c@{\hskip8pt}c@{\hskip8pt}c@{\hskip8pt}c@{\hskip8pt}c@{\hskip8pt}c@{\hskip8pt}c}
                \hline
                Model Name           & P       & F1 score   & IoU   & OA     \\
                \hline
                FCN                  & 0.7045        & 0.7242     & 0.5676  & 0.9932 \\ 
                FCN-CG             & 0.7240        & 0.7362     & 0.5826 & 0.9935 \\
                DeepLabv3            & 0.7129        & 0.7337     & 0.5794 & 0.9935 \\
                DeepLabv3-CG       & \textbf{0.7523}        & \textbf{0.7508}     & \textbf{0.6010} & \textbf{0.9937} \\
                \hline
            \end{tabular}
        \end{table}

\subsection{Qualitative Evaluation}

In addition to the quantitative evaluation, we visualize several segmentation results in Fig. \ref{fig:subgisV} and \ref{fig:subgisF}. Pixel-based true positives, false positives, and false negatives are presented in green, red, and blue, respectively. 

Fig. \ref{fig:subgisV} shows results of models using sensor-visible footprint segments. 
We can observe a general improvement in quality from FCN/DeepLabv3 to FCN-CG/DeepLabv3-CG, especially for buildings in column $b$, $c$, and $g$.
For buildings with simple structures (e.g., the building in column $a$), all models are able to offer satisfactory segmentation results, while for those with complicated shapes (see column $e$), large under-segmentation areas (cf. red pixels) can be seen in predicted building masks. Besides, the utilization of the proposed CG module can effectively reduce over-segmentation in final predictions. 

Fig. \ref{fig:subgisF} presents results of models using complete footprints. They indicate that our CG module can ease both over-segmentation (cf. blue pixels in column $b$) and under-segmentation (cf. red pixels in column $e$) problems to a considerable extent.
Moreover, examples in the third row, column $f$ and the fifth row, column $f$ show that the connectivity of segmentation results are disrupted (cf. green pixels), while the integration of the CG module can alleviate such a problem. 
A similar phenomenon can also be seen in column $d$ and $g$ that exploiting the CG module can enhance the connectivity of predictions.
In summary, the proposed CG module effectively improves segmentation results.

\subsection{Comparison of Complete Building Footprints and Sensor-visible Footprint Segments}

From Table \ref{tab:gisV} and \ref{tab:gisF}, we can see that models trained with complete building footprints surpass those trained with sensor-visible footprint segments. For instance, DeepLabv3-CG trained on complete footprints improves the F1 score and IoU by 4.40\% and 5.45\%, respectively, compared to that learned with sensor-visible segments. 

Fig. \ref{fig:patch} provides segmentation results of two patches using two versions of footprint masks, and different buildings are marked in different colors (50\% transparency).  
Note that individual building masks are predicted separately, and then masks of buildings in the same patch are plotted together to visualize the overlapping areas.
Here, patch 1 presents a simple scenario, in which buildings are isolated and show clear signatures in the SAR image. In this case, all models can obtain good segmentation results.
Patch 2 shows a fairly complicated scene, where two consecutive buildings exist in the center (cf. buildings in cyan and blue), and SAR signatures are unclear. 
Although all networks can still successfully segment isolated buildings, the two overlapped buildings are not correctly segmented by models trained with sensor-visible footprint segments (see the third row of Fig.~\ref{fig:patch}). 
This is because the mask of sensor-visible footprint segments for the building on the left contains only one edge, which does not provide adequate information.
Moreover, we notice that the overlapping region between these two buildings can only be well identified by models trained with complete building footprints.

Overall, 
these results suggest that complete building footprints are more befitting for segmentation of individual buildings than sensor-visible footprint segments. 
This may be because the former delivers more information, especially for low-rise buildings.

\subsection{Can CG-Net work with inaccurate GIS data?}

So far, building footprints used in our experiments are highly accurate as they are acquired from official GIS data. However, most openly available GIS data, such as OpenSteetMap (OSM), often contain positioning errors. To test the performance of CG-Net in such cases, we conduct supplementary experiments on training our CG-Net with inaccurate building footprints, and discuss the impact of positioning errors in GIS data.

\begin{figure}
        \centering
        \includegraphics[width=.6\columnwidth]{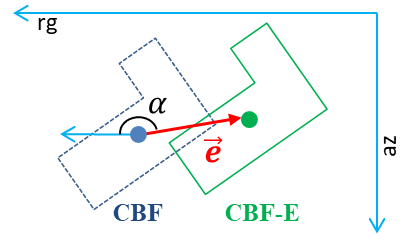}
        \caption{Illustration of generating building footprints with positioning errors. Positioning error {$\protect\overrightarrow{e}$} is added to building footprint CBF, resulting in CBF-E. rg and az denote the range direction and azimuth direction, respectively. $\alpha$ is the angle between {$\protect\overrightarrow{e}$} and rg. } 
        \label{fig:ftp_e} 
    \end{figure}

First, we generate inaccurate CBF, termed as CBF-E, by injecting positioning errors. As illustrated in Figure \ref{fig:ftp_e}, $\protect\overrightarrow{e}$ is the added positioning error, and $\alpha$ is the angle between $\protect\overrightarrow{e}$ and the range direction. 
According to the quality assessment study of OSM in \cite{fan2014quality}, the average offset of building footprints is 4.13 m with the standard deviation of 1.71 m. 
Therefore we consider the positioning error as a variable whose magnitude is Gaussian distributed, i.e., $|\protect\overrightarrow{e}| \sim \mathcal{N}(\mu=4.13, \sigma^2=1.71^2)$.   
Since the offset may point to different directions, we assume the direction of $\protect\overrightarrow{e}$ is uniformly distributed, i.e., $\alpha$  is uniformaly distributed in the range of $[0^{\circ},360^{\circ})$. 
For simplicity, let $\alpha$ be discrete: $\alpha \sim$ DiscreteUniform$(0^{\circ},359^{\circ})$. 
Note that this is the most difficult case that all footprints contain positioning errors. 
   
Then, we train DeepLabv3-CG using CBF-E and SAR patches, and test the trained network with a clean test set. DeepLabv3-CG is chosen because it performs best among all the networks. The parameter settings of the network remain the same as previous experiments, as described in Section \ref{sec:train_details}.

\begin{figure}[!]
    \centering
    \settoheight{\tempdima}{\includegraphics[width=.15\textwidth]{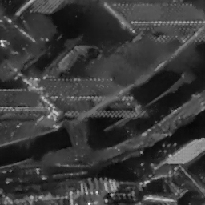}}%
    \begin{tabularx}{0.99\linewidth}{c@{\hskip2pt}c@{\hskip2pt}c@{\hskip2pt}c@{\hskip2pt}c@{\hskip2pt}}
    \\\vspace{-0.4cm}
    \textbf{\hspace{0.3cm}\footnotesize{a}} &\textbf{\footnotesize{b}}& \textbf{\footnotesize{c}}
    \\\vspace{-0.4cm}        \rowname{\hspace{-0.3cm}\scriptsize{\textcolor{white}{y}}} 
    \rowname{\footnotesize{SAR image}} 
    \hspace{-0.06cm}\subfloat{{\includegraphics[width=.13\textwidth]{figs/secF/06101_0.png}}}&
    \subfloat{{\includegraphics[width=.13\textwidth]{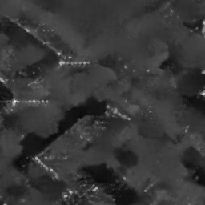}}}& 
    \subfloat{{\includegraphics[width=.13\textwidth]{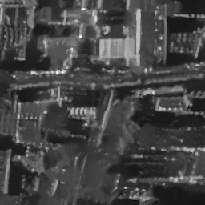}}}&
    \\\vspace{-0.4cm}
    \rowname{\hspace{-0.3cm}\scriptsize{\textcolor{white}{y}}} 
    \rowname{\hspace{-0.3cm}\footnotesize{CBF}} 
    \subfloat{{\includegraphics[width=.13\textwidth]{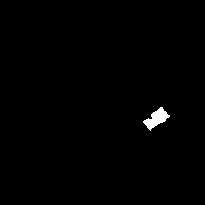}}}&
    \subfloat{{\includegraphics[width=.13\textwidth]{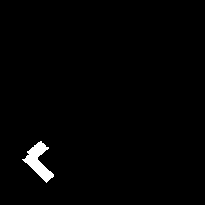}}}& 
    \subfloat{{\includegraphics[width=.13\textwidth]{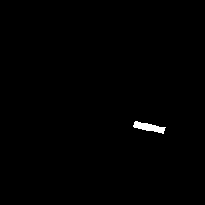}}}&
    \\\vspace{-0.4cm}
    \rowname{\hspace{-0.3cm}\scriptsize{DeepLabv3-CG}}
    \rowname{\hspace{-0.3cm}\scriptsize{trained on CBF}}
    \subfloat{{\includegraphics[width=.13\textwidth]{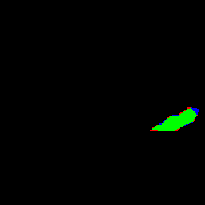}}}&
    \subfloat{{\includegraphics[width=.13\textwidth]{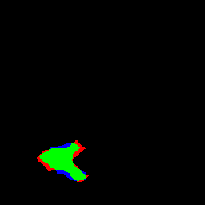}}}& 
    \subfloat{{\includegraphics[width=.13\textwidth]{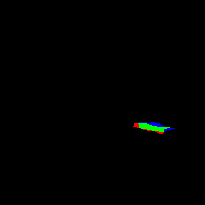}}}&
    \\\vspace{-0.4cm}
    \rowname{\hspace{-0.3cm}\scriptsize{DeepLabv3-CG}}
    \rowname{\hspace{-0.3cm}\scriptsize{trained on CBF-E}}
    \subfloat{{\includegraphics[width=.13\textwidth]{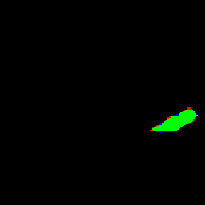}}}&
    \subfloat{{\includegraphics[width=.13\textwidth]{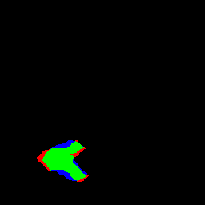}}}& 
    \subfloat{{\includegraphics[width=.13\textwidth]{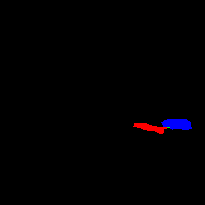}}}&
    \end{tabularx}
    \caption{Examples of segmentation results of networks trained using complete building footprints (abbreviated as CBF) and networks trained using building footprints with positioning errors (abbreviated as CBF-E). Pixel-based true positives, false positives, and false negatives are marked in green, red, and blue, respectively.}
    \label{fig:cbfe_gt}
\end{figure}

The results are listed in Table \ref{tab:cbfe}.  As can be seen, comparing to results using CBF, the precision of the network trained on CBF-E is decreased by 3.02\%, the F1 score is reduced by 3.62\%, and the IoU is decreased by 4.5\%. However it still gives competent segmentation results.
For visual comparison, Figure \ref{fig:cbfe_gt} shows results of DeepLabv3-CG trained with CBF-E and CBF. For the building in column c, DeepLabv3-CG trained with CBF performs much better than that with CBF-E. However the predictions for buildings in column a and b are visually very similar. Moreover, we observed that predictions from DeepLabv3-CG trained on CBF-E are satisfactory for most buildings. 
        \begin{table}[!]
            \centering
            \caption{Numerical results of DeepLabv3-CG trained using CBF and CBF-E.}
            \begin{tabular}{@{\hskip8pt}c@{\hskip8pt}c@{\hskip8pt}c@{\hskip8pt}c@{\hskip8pt}c@{\hskip8pt}c@{\hskip8pt}c@{\hskip8pt}c@{\hskip8pt}c@{\hskip8pt}c}
                \hline
                GIS data used for training           & P       & F1 score   & IoU   & OA     \\
                \hline
                CBF       & {0.7523}        & {0.7508}     & {0.6010} & {0.9937} \\
                CBF-E            & 0.7221        & 0.7146     & 0.5560 & 0.9927 \\
                \hline
            \end{tabular}
            \label{tab:cbfe}
        \end{table}

The experiments show that although weakened by positioning errors in GIS data, the proposed CG-Net is robust even in the most difficult case. This finding suggests that the large amount of existing open sourced GIS data, such as OSM, can be exploited for segmenting individual buildings in SAR images.

\vspace{12pt}

\section{Further application: reconstruction of LoD1 building models from a SAR image}\label{sec:app}

          \begin{figure}[t!]
            \centering
            \includegraphics[width=0.37\textwidth]{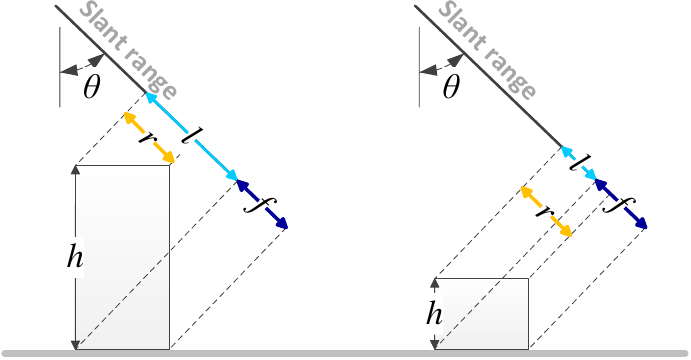}
            \caption{The projection geometry of two flat-roof buildings in a slant-range SAR image. $\theta$ is the incidence angle. $h$ is the building height. 
            \textit{l}, \textit{r}, and \textit{f} denote the length of layover, roof, and footprint areas in a slant-range SAR image, respectively. 
             }
            \label{fig:hl_hm}
        \end{figure}

\begin{figure*}[!]
    \centering
    \includegraphics[trim=1.5cm 0cm 0cm 1.5cm, clip=true,width= \textwidth]{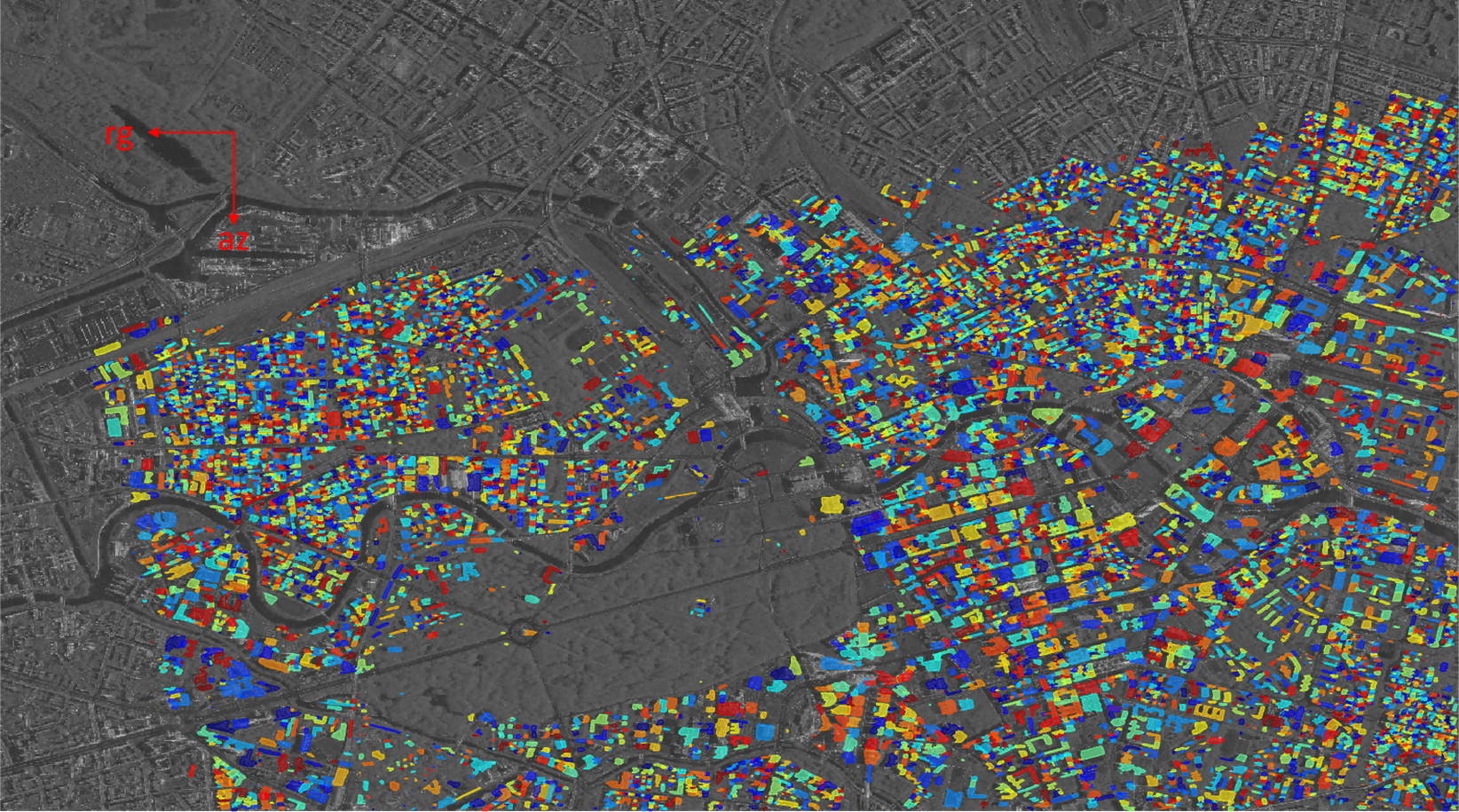}
    \caption{Segmentation results in the study area obtained by DeepLabv3-CG. The building segments are plotted with different colors translucently for visualising the layover areas between buildings. rg and az denote the range direction and azimuth direction, respectively.}
    \label{fig:bigarea}
\end{figure*}

    Building models can be created at different levels-of-detail (LoD). 
    According to the terminology of CityGML \cite{kolbe2005CityGML}, LoD1 models represent buildings as blocks with flat roof structures and can be reconstructed by extruding footprints with building heights. Here, we regard the average roof height as the building height\footnote{http://en.wiki.quality.sig3d.org/index.php/Modeling\_Guide\_for\_3D\_Objects \_-\_Part\_2:\_Modeling\_of\_Buildings\_(LoD1,\_LoD2,\_LoD3)}. 
    In this section, we demonstrate the process of reconstructing LoD1 models using our predicted individual building masks.

    Fig. \ref{fig:hl_hm} illustrates the projection geometry of two flat-roof buildings in a constant azimuth profile of a SAR image.  
    $\theta$ is the incidence angle. \textit{l}, \textit{r}, and \textit{f} denote the length of layover, roof, and footprint areas in the slant-range SAR image, respectively. 
    Notably, the building region in the SAR image contains both the layover and the roof areas. The layover area coincides with the building region when the building height \textit{h} is large, e.g., the case in Fig. \ref{fig:hl_hm} (left), and it is covered by the building region when $h$ is small, e.g., the case in Fig. \ref{fig:hl_hm} (right). 
    In both cases, the layover area can be calculated by subtracting the footprint from the building region. Therefore, \textit{l} is estimated to be the length of the layover area in the slant-range direction, and \textit{h} can be computed with the following equation: 
    \begin{equation}
    \label{eq:height}
        h = l/cos \theta.
    \end{equation}

\begin{figure}[!]
            \centering
            \includegraphics[width= 0.5\textwidth]{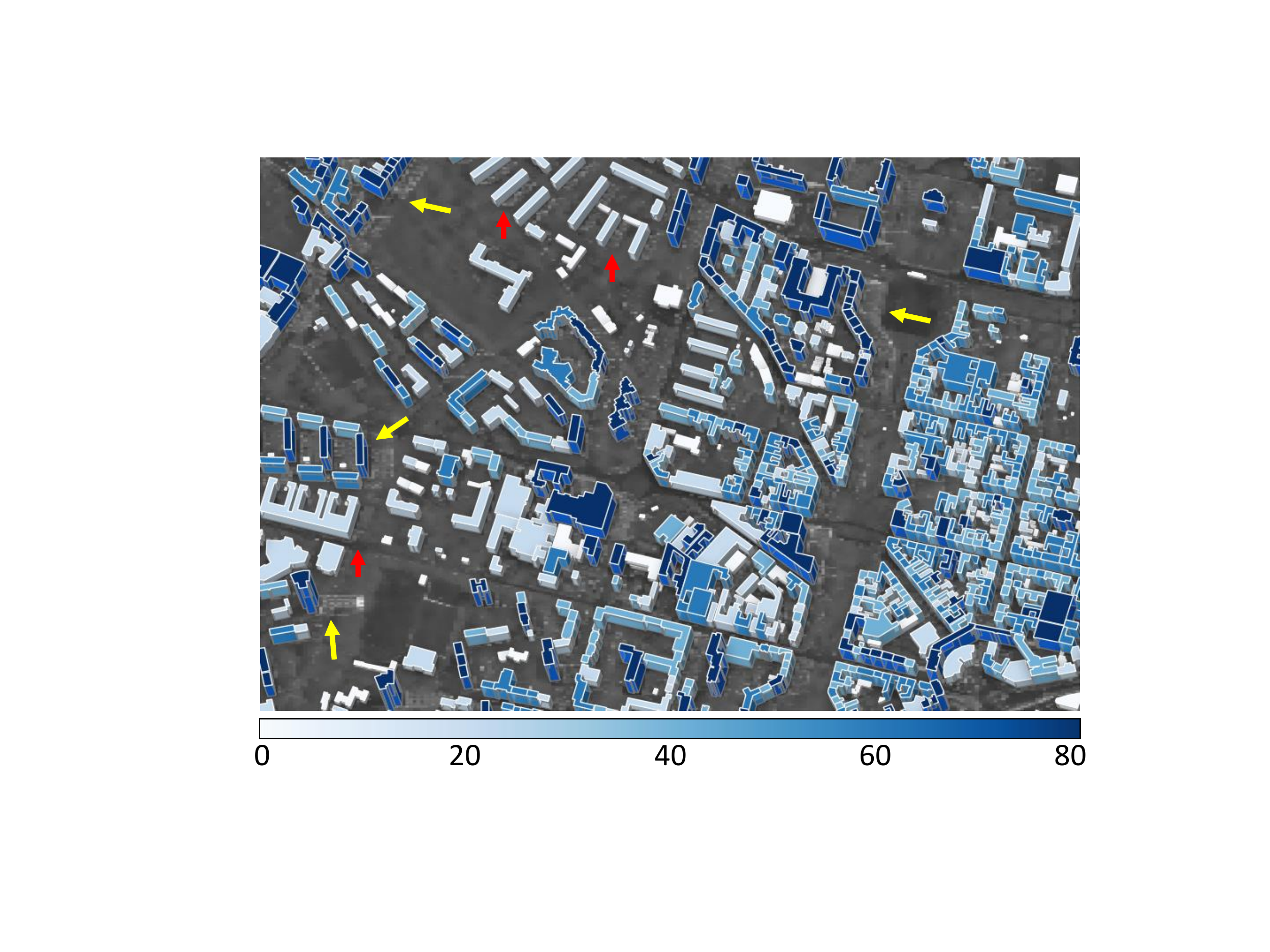}
            \caption{Example LoD1 building models in the study area superimposed on the SAR image. 
            Layover areas of some buildings are visible, as pointed by the yellow and red arrows. Building heights are color-coded.}
            \label{fig:lod1}
        \end{figure}

\begin{figure}[!]
            \centering
            \includegraphics[width= 0.4\textwidth]{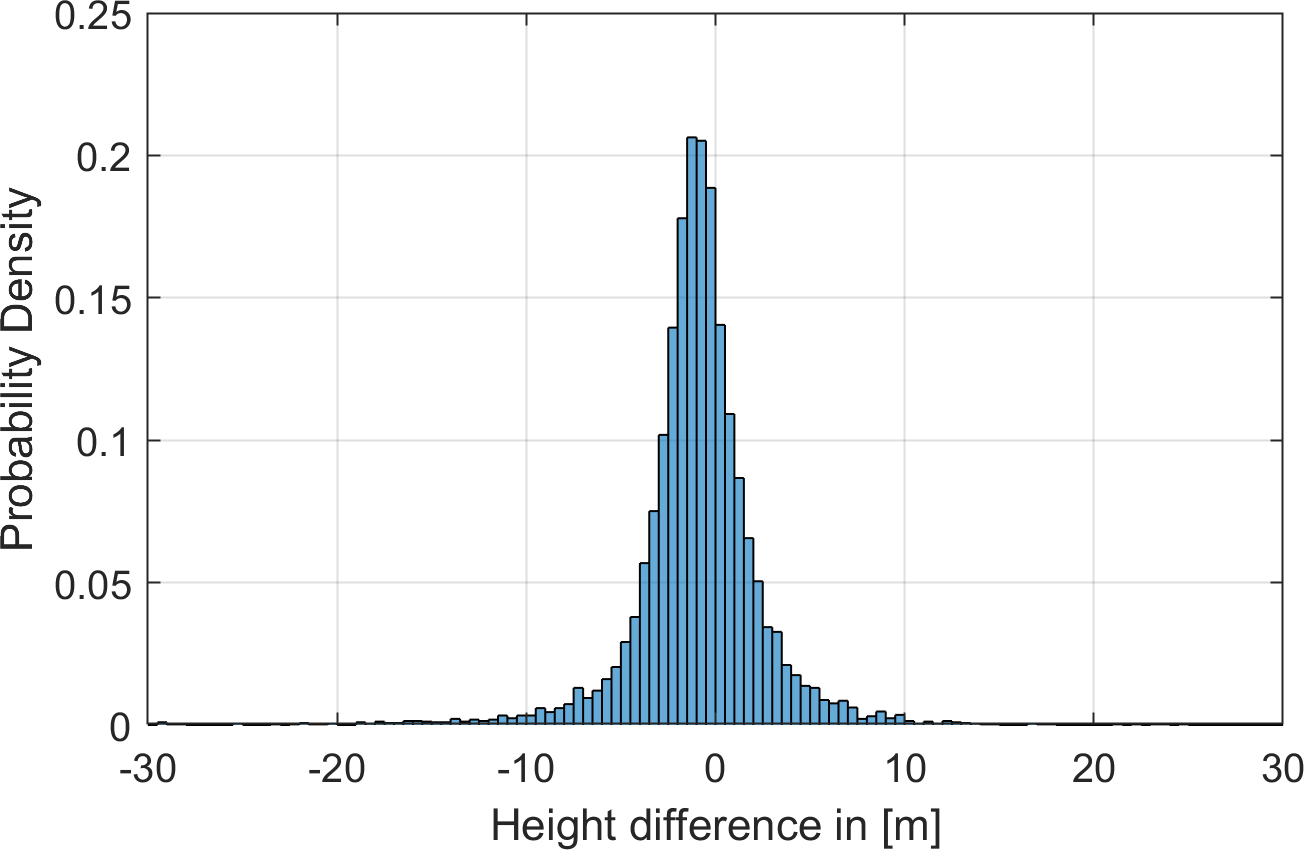}
            \caption{Histogram of building height errors in the study area. }
            \label{fig:he_Hist}
        \end{figure}

    From the predicted individual building masks (cf. Fig. \ref{fig:bigarea}), we calculate building heights with Eq. (\ref{eq:height}). Afterwards,
    LoD1 building models are created by extruding building footprints with obtained heights.  
    Fig. \ref{fig:lod1} presents example LoD1 models superimposed on the SAR image in the study area. It can be observed that buildings with large $l$ (pointed by yellow arrows) are predicted as high-rise, while those with small $l$ (pointed by red arrows) are reconstructed as low-rise buildings. 
    This is in line with the reality.
    We further evaluate the estimated height against the mean height from the accurate DEM for each building. 
    The mean height error we achieve in the study site is 2.39 m. 
    The histogram of height errors is shown in Fig. \ref{fig:he_Hist}.

\section{Conclusion}\label{sec:conclude}

In this paper, we propose a conditional GIS-aware network (CG-Net) to segment individual buildings from a large-scale VHR SAR image. 
We also propose an approach for generating ground truth annotations of buildings using a high resolution DEM. 
The proposed method is evaluated in Berlin area, using a high resolution spotlight TerraSAR-X image and building footprints obtained from GIS data. Both qualitative and quantitative results demonstrate the effectiveness of the proposed CG-module. 
Compared to competitors, DeepLabv3-CG achieves the best F1 score of 75.08\%. 
In addition, we compare two building footprint representations, namely complete building footprints and sensor-visible footprint segments. Experimental results suggest that the use of complete building footprints leads to better results. 
Further experiments of training the networks using inaccurate GIS data suggest that CG-Net is robust in presence of positioning errors in GIS data.
Additionally, we demonstrate an application of our results, i.e., LoD1 building model reconstruction. 
In the future, 
we are interested in applying the proposed data generation workflow to areas of various urban morphologies and using our CG-Net to reconstruct LoD1 building models from TerraSAR-X and TanDEM-X stripmap images.

\section*{Acknowledgment}

The authors would like to thank Dr. H. Hirschmüller of DLR-RM for providing the optical DEM. 

\bibliographystyle{IEEEtran}
\bibliography{IEEEabrv,mainbib}

\end{document}